\documentclass[a4paper,12pt]{JHEP3}
\usepackage{amsmath,amsfonts,epsfig}
\def\th#1.#2{\theta_1 (z_{#1} - z_{#2})}
\def\thn#1.#2.#3.#4{\theta_{\nu} (\frac{#1 z_{1} #2 z_{2} #3 z_3 #4 z_4}{2} )}
\def\us#1.#2.#3.#4{(u_{#1}\wedge u_{#2}) (u_{#3} \wedge u_{#4})} 
\def\bra{\langle}
\def\ket{\rangle}
\def\d{\mathrm{d}}
\def\bb#1{\mathbb{#1}}

\def\ad#1{\dot{\alpha}_{#1}}
\def\lk{\lambda^{\kappa}}
\def\nk{\nu^{\kappa}}
\def\tr{\mathrm{tr}}
\def\ap{\alpha^{\prime}}

\numberwithin{equation}{section}

\preprint{IPPP/05/75\\ DCPT/05/150\\hep-th/0512072}

\author{Steven A. Abel and Mark D. Goodsell\\ Centre for Particle Theory, University of Durham, Durham, DH1 3LE, UK\\
 E-mail: \email{s.a.abel@durham.ac.uk}, \email{m.d.goodsell@durham.ac.uk}}

\title{Intersecting Brane Worlds at One Loop}

\abstract{
We develop techniques for one-loop diagrams on 
intersecting branes. The one-loop propagator of chiral intersection 
states on D6 branes is
calculated exactly and its finiteness is shown to be guaranteed by RR
tadpole cancellation. The result is used to demonstrate
the expected softening of power law running of Yukawa couplings
at the string scale. We also
develop methods to calculate arbitrary $N$-point functions at
one-loop, including those without gauge bosons in the loop. These
techniques are also applicable to heterotic orbifold models.}


\begin{document}

\tableofcontents

\section{Introduction}

Intersecting Brane Worlds in type IIA string theory, consisting of D6
branes wrapping $\bb{R}_4 \times \bb{T}_2 \times \bb{T}_2 \times
\bb{T}_2$, have proven popular for constructing toy models with many
attractive features (see
e.g. \cite{Blumenhagen:1999ev,Cvetic:Models,Kiritsis:2003mc,
Uranga:2003pz,Cvetic:2003xs,Lust:2004ks,Kokorelis:2004dc,Blumenhagen:2004vz,Uranga:2005wn,Blumenhagen:2005tn}
or \cite{Blumenhagen:2005mu} for a review). One of the interesting
features of these models is that, unlike heterotic orbifolds, wrapping
D-branes leads to a number of geometric moduli that can be easily
adjusted in order to set gauge couplings. In addition the localization
of matter fields at D-brane intersections may have implications for a
number of phenomenological questions, most notably Yukawa coupling
hierarchies \cite{Cremades:yukawa}. Indeed these can be understood by
having the matter fields in the coupling located at different
intersections, with the resulting coupling being suppressed by the
classical world-sheet instanton action (the minimal world-sheet area
in other words).  This is appealing because it suggests a possible
geometric explanation of Yukawa hierarchies in terms of compactification geometry. This realization led to subsequent work to examine
tree-level couplings in greater detail
\cite{Cvetic:Yukawa,Abel:2003vv,Owen,Cremades:2004wa}. 
For the Yukawa couplings exact calculations have confirmed that the
normalization (roughly speaking the quantum prefactor in the
amplitude) can be attributed to the K\"ahler potential of matter fields, with the
coupling in the supergravity basis being essentially unity (or zero).

Currently the K\"ahler potential in {\em chiral matter} superfields is known 
to only quadratic terms and only at tree level \cite{Lust:scat,Bertolini:2005qh} 
(one loop corrections to the moduli
sectors of K\"ahler potentials in IIB models have been calculated in
\cite{Berg:2005ja}).
For a multitude of phenomenological reasons it is something we would like to 
understand better, especially its quantum corrections. In this 
paper we go a step further in this direction with an analysis of 
interactions of chiral matter fields at one-loop.

\begin{figure}
\begin{center}
\epsfig{file=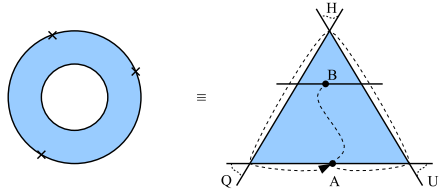,height=1.0in}
\caption{An annulus contribution to 3 point functions.}\label{fig:annulus}\end{center}
\end{figure}

Figure \ref{fig:annulus} shows the physical principle of calculating a
one-loop annulus for the example of 
a 3 point coupling, discussed in ref.\cite{Abel:2003yh}. 
Take a string stretched between two branes as shown and keep one end (B) fixed on a particular brane, 
whilst the opposing end (A) sweeps out a triangle (or an $N$-sided polygon for 
$N$-point functions). Chiral states 
are deposited at each vertex as the endpoint A switches from one brane to the next.
As the branes are at angles and hence the open string endpoints free to move 
in different directions, the states have ``twisted'' boundary conditions, 
reflected in the vertex operators by the inclusion of so-called twist operators. 
Working out the CFT of these objects is usually the most arduous part of 
calculations on intersecting branes. The
corresponding worldsheet diagram is then the annulus with 3 ($N$) vertex operators. 
In the presence of orientifolds there will be M\"obius strip diagrams as well.
There is no constraint on the relative positioning of the B brane (although the usual 
rule that the action goes as the square of the brane
separation will continue to be obeyed), and it may be one of the other branes. 
(The case that brane B is at angles to the three other branes was not 
previously possible to calculate: we shall treat it in detail in this paper.)

Finding the K\"ahler potential means extracting two point functions, but because 
the theory is defined on-shell we have to take an indirect approach;
complete answers can be 
obtained only by factorising down from at least 4 point functions. To do 
this we first develop the general formalism for $N$-point
functions in full. However it would be extremely tedious if we had to 
factorise down the full amplitude including the classical 
instanton action for every 
$N$-point diagram. Fortunately the OPE rules for the chiral states at
intersections allow a short cut: first we use the fact that 
the superpotential is protected by
the non-renormalization theorem which has a stringy
incarnation derived explicitly in \cite{Ben}. 
In supersymmetric theories, the interesting diagrams are therefore the field 
renormalization diagrams which we can get in the field 
theory limit where two vertices come together. 
A consistent procedure therefore is to use the OPE rules 
to factor pairs of {\em external} states onto a single state 
times the appropriate tree-level Yukawa coupling. In this way one 
extracts the {\em off-shell} two-point function.

This procedure allows us to consider various aspects of
one-loop processes.  For example, in $N=1$ supergravity
the non-renormalization theorem does not protect the K\"ahler
potential, and only particular forms of K\"ahler potential
(essentially those with $\log\det(K_{ij})=0$ where $K_{ij}$ is the
K\"ahler metric) do not have quadratic divergences at one-loop order
and higher (for a recent discussion see \cite{Gaillard:2005cw}). It is
natural to wonder therefore how string theory ensures that such divergences are
absent. Here we show that the conditions for cancellation of the
divergences is identical to the Ramond-Ramond tadpole cancellation
condition (essentially because in performing the calculation one
factors down onto the twisted partition function). (At the level of
the effective field theory there are two well known forms of
tree-level K\"ahler metric that are consistent with the one-loop
cancellation of divergences, the ``Heisenberg'' logarithmic form, and
the ``canonical'' quadratic form). As a second application, we
consider the subsequent determination of field renormalization; 
we find agreement with the power law running that one
deduces from the effective field theory. However we also see that
as we approach the string scale the power law running dies away 
as the string theory tames the UV divergences. 

\section{One-Loop Scalar Propagator}

\subsection{Four-Point Amplitudes}

String theory is defined only on-shell, so in order to calculate
the energy dependence of physical couplings we must calculate physical
diagrams and probe their behaviour as the external momenta are
varied. To obtain the running of Yukawa couplings in
intersecting brane models, we should in principle consider the 
full four-point amplitudes,
since they are the simplest diagrams with non-trivial Mandelstam
variables. However here we shall take the more efficient approach
outlined in the introduction: 
due to the existence of a consistent off-shell
extension of these amplitudes, it is possible to calculate three-point
diagrams to obtain the Yukawa couplings. This was done in ref.\cite{Ben} for
the amplitude in certain limits (albeit with some inevitable ambiguities). 
There it was shown that for
supersymmetric amplitudes the nonrenormalization theorem 
held as expected, and the low-energy behaviour is entirely dominated by
the renormalization of propagators. The latter {\em can} be consistently 
calculated in full by factorising down from the full four-point functions, 
and this is what we shall do here.

We shall focus on four-fermion amplitudes, since they factorise onto
the lower-vertex amplitudes of interest - the yukawa renormalisation
amplitude and the two scalar amplitude. In intersecting brane models, the
allowed amplitudes are constrained by the necessity of the total
boundary rotation being an integer; supersymmetry then dictates the
allowed chirality of the vertex operators via the GSO projection. In
the case of $N=1$ supersymmetry, the requirement of integer rotation
results in two pairs of opposite chirality fermions being required. In
the case of $N=2$ or 4 supersymmetry, we are also allowed to have four
fermions of the same chirality. The correlator for the non-compact
dimensions in the case of $N=1$ supersymmetry was calculated in
ref.\cite{AtickDixonSen}, in order to legitimise their calculations on
orbifolds. We have checked that the $N=2$-relevant amplitude gives the
same limit upon factorisation. Thus, we can simply use their result to
justify using the OPE behaviour of the vertex operators in order 
to obtain the
low-energy limit of the four-point function, and we can proceed with
the calculation of the scalar propagator to obtain the corrections to
the Yukawa couplings. In the next subsection we outline some of the
technology required, and in the ensuing subsections we extract the
information about the running of the couplings.

\subsection{Vertex Operators}

To begin, we review the technology, assembled in ref.\cite{Ben}, for the
calculation of loop amplitudes involving massless chiral
superfields. The appropriate vertex operator to use for incoming
states at an intersection depends on the angles in each
subtorus. There are several possible conditions for supersymmetry (we
will focus on N=1 supersymmetric models) of which we will use the most
straightforward: for \emph{intersection} angles $\theta^{\kappa}$
(where $\kappa$ runs over the complex compact dimensions 1 to 3) we
have $\sum_{\kappa} \theta^{\kappa}=1 \ \mathrm{or}\ 2$. We have two
possible conditions because each intersection supports one chiral and
one antichiral superfield, with complimentary angles. With this
condition the GSO projection correlates the chirality of the fermions
with that of the rotation, and we obtain vertex operators as in
\cite{Cvetic:Yukawa}:

\begin{eqnarray}
V_{-1}^{(ab)} (k,\theta^{\kappa},z) &=& \lambda^{(ab)} e^{-\phi} e^{ik \cdot X} \prod_{\kappa} e^{i\theta^{\kappa} H^{\kappa}} \sigma_{\theta^{\kappa}}^{(ab)} (z) \\
V_{-1/2}^{(ab)} (u,k,\theta^{\kappa},z) &=& \lambda^{(ab)} e^{-\frac{\phi}{2}} u^{\dot{\alpha}} \tilde{S}_{\dot{\alpha}} e^{ik\cdot X} \prod_{\kappa=1}^{3} e^{i (\theta^{\kappa}-1/2) H^{\kappa}} \sigma_{\theta^{\kappa}}^{(ab)} (z) \nonumber
\end{eqnarray}
for $\sum_{\kappa} \theta^{\kappa} = 1$, and for the ``antiparticle'':
\begin{eqnarray}
V_{-1}^{(ab)} (k,1-\theta^{\kappa},z) &=& \lambda^{(ab)} e^{-\phi} e^{ik \cdot X} \prod_{\kappa} e^{-i\theta^{\kappa} H^{\kappa}} \sigma_{1-\theta^{\kappa}}^{(ab)} (z) \\
V_{-1/2}^{(ab)} (u,k,1-\theta^{\kappa},z) &=& \lambda^{(ab)} e^{-\frac{\phi}{2}} u^{\beta} S_{\beta} e^{ik\cdot X} \prod_{\kappa=1}^{3} e^{-i (\theta^{\kappa}-1/2) H^{\kappa}} \sigma_{1-\theta^{\kappa}}^{(ab)} (z) \nonumber 
\end{eqnarray}
The worldsheet fermions have been written in bosonised form, where the
coefficient $\alpha$ in $e^{i\alpha H}$ shall be referred to as the
``H-charge''. The spacetime Weyl spinor fields are the left-handed
$\tilde{S}_{\alpha} = e^{\pm\frac{1}{2}(\mathcal{H}_0 -
\mathcal{H}_1)}$ and right handed $S_{\beta} =
e^{\pm\frac{1}{2}(\mathcal{H}_0 + \mathcal{H}_1)}$. The operators
$\sigma_\theta^{(ab)}$ are boundary-changing operators (here between
branes $a$ and $b$), whose properties are discussed in Appendix
\ref{CFT}.

$\lambda^{(ab)}$ is the appropriate Chan-Paton factor for the
vertex. We shall not require the specific properties of these, but in
the amplitudes we consider they are accompanied by model-dependent
matrices $\gamma^{a}_{i}$ which encode the orientifold
projections. These matrices have been described for many models
(e.g. \cite{Blumenhagen:1999ev,Forste:2000hx,Cvetic:Models}, but we
will only need the results given in \cite{Lust:Gauge} for
$\mathbb{Z}_N$ or $\mathbb{Z}_N \times \mathbb{Z}_M$ orientifolds:
\begin{eqnarray}
\gamma^{a}_1 &=& \mathbf{1}_{N_a} \nonumber \\
\tr \gamma^a_{\hat{\theta}^{N/2}} = \tr \gamma^a_{\hat{\omega}^{M/2}} = \tr \gamma^a_{\hat{\theta}^{N/2}\hat{\omega}^{M/2}}&=& 0\nonumber \\
(\gamma^{\Omega R \hat{\Theta}^{k,l}a}_{\Omega R \hat{\Theta}^{k,l}})^* \gamma^{a}_{\Omega R \hat{\Theta}^{k,l}} &=& \rho_{\Omega R \hat{\Theta}^{k,l}} \mathbf{1}_{N_a}
\end{eqnarray}
where $\hat{\theta}^{k}$ is a $\mathbb{Z}_N$ twist, $\hat{\omega}^{k}$
is also present in $\mathbb{Z}_N \times \mathbb{Z}_M$, and
$\rho_{\Omega R \hat{\theta}^k} = \pm 1$. $k$($l$) runs from $0$ to
$N-1$ ($M-1$) for $\hat{\theta}^{k}$ ($\hat{\omega}^{l}$), where
$\hat{\theta}^0=\hat{\omega}^0 = 1$. In the last expression we have
used the notation $\hat{\Theta}^{k,l} =
\hat{\theta}^{k}\hat{\omega}^{l}$, and so for example in the
$\mathbb{Z}_2 \times \mathbb{Z}_2$ orientifold we have $\rho_{\Omega R
\hat{\Theta}^{1,0}} = \rho_{\Omega R \hat{\Theta}^{0,1}}=\rho_{\Omega
R \hat{\Theta}^{1,1}}= -1$ and $\rho_{\Omega R \hat{\Theta}^{0,0}}=1$.
The  massless four-fermion amplitude that we shall consider is given by
\begin{multline}
\mathcal{A}_4^1 = \int_0^{\infty} \d t\prod_{i=1}^4 \int \d z_i \\
\bra V_{+1/2}^{(ca)}(u^{\alpha_4}_4,k_4,z_4) V_{-1/2}^{(ab)} (u^{\alpha_1}_1,k_1,z_1) V_{-1/2}^{(ba^{\parallel})}(u^{\ad2}_2,k_2,z_2)V_{+1/2}^{(a^{\parallel}c)} (u^{\ad3}_3,k_3,z_3)  \ket_{cc}
\end{multline}
where brane $a^{\parallel}$ is parallel to brane $a$ (or $a$ itself in the simplest case).

\subsection{Field Theory Behaviour}

The behaviour of the amplitude in the
field-theory limit is found by considering the momenta to be small. When we
do this we find that, due to the OPEs of the vertex operators, the
amplitude is dominated by poles where the operators are contracted
together to leave a scalar propagator. From the previous section and
the results of Appendix \ref{CFT}, we find that two fermion vertices 
factorise as 
\begin{multline}
V_{-1/2}^{(bd)} (u^{\ad2},k_2,1-\nu^{\kappa},z_2) V_{+1/2}^{(dc)} (u^{\ad3},k_3,,1-\lambda^{\kappa},z_3)\\
 \sim (u_2 u_3)(z_2 - z_3)^{2\alpha^{\prime} k_2\cdot k_3 - \sum_{\kappa}\frac{\nu^{\kappa} + \lambda^{\kappa}}{2}} V_{0} (k_2 + k_3,z_2) g_o\prod_{\kappa} C^{(bdc)1-\nu^{\kappa}-\lambda^{\kappa}}_{\nu^{\kappa}, \lambda^{\kappa}}
\end{multline}
for $\sum_{\kappa}\nu^{\kappa} = \sum_{\kappa}\lambda^{\kappa} = 1$,
and $C^{(bdc)1-\nu^{\kappa}-\lambda^{\kappa}}_{\nu^{\kappa},
\lambda^{\kappa}}$ are the OPE coefficients, given in equation
(\ref{abcOPE}). The calculation involves a factorisation of
the tree-level four-point function on first the gauge exchange and 
then the Higgs exchange, and a comparison with the field theory 
result \cite{Cvetic:Yukawa}. 

Note that for consistency 
the classical instanton contribution should be included in the 
OPE coefficients as well as the quantum part. This means
that as we go on to consider higher order loop diagrams 
the tree-level Yukawa couplings (including classical contributions) 
should appear in the relevant field theory limits. 
In the factorisation limit these two fermions yield the
required pole of order one around $z_2$; we
will obtain a similar pole for the other two fermion vertices. If we
were to then integrate the amplitude over $z_3$ we would obtain a
propagator $\frac{1}{2\ap k_2 \cdot k_3}$ preceding a three-point
amplitude, and performing the integration over $z_4$, say, we would
obtain another propagator $\frac{1}{2\ap k_1 \cdot k_4}$
($=\frac{1}{2\ap k_2 \cdot k_3}$), and have reduced the amplitude to a
two-scalar amplitude. 

\begin{figure}
\begin{center}
\epsfig{file=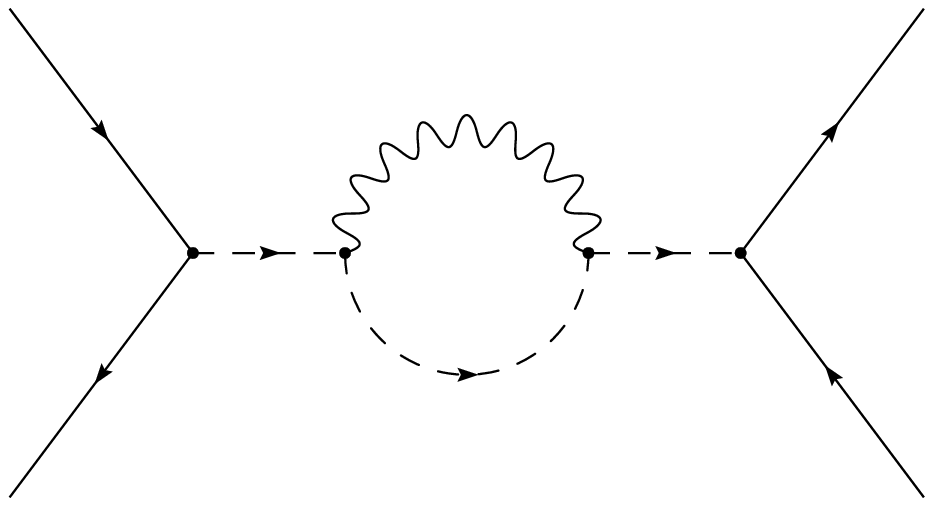,height=1.0in}$\qquad$ \epsfig{file=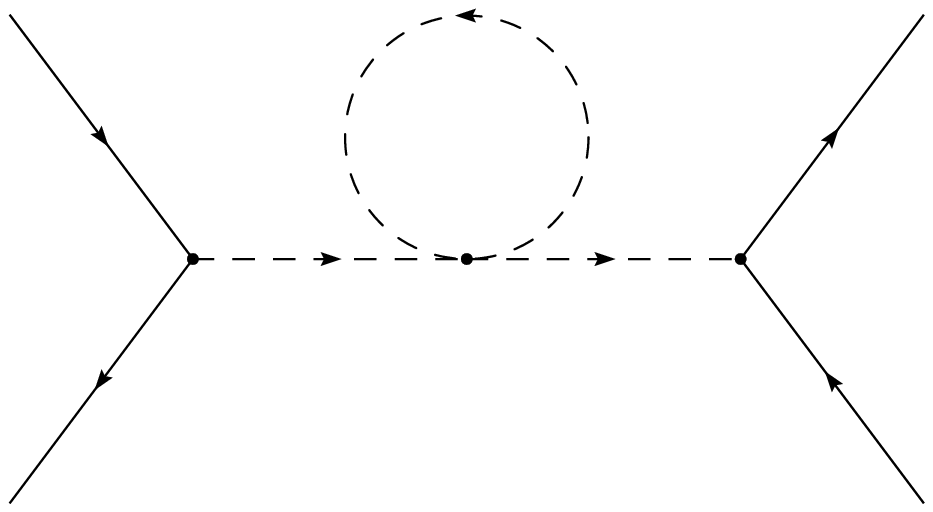,height=1.0in}
\caption{Feynman diagrams in the field theory equivalent to our limit; we have factorised onto the scalar propagator, and consider only gauge and self-couplings.}\label{fig:feynman}\end{center}
\end{figure}

In this manner the four-point function can be factorised onto the 
two point function and reduces to 
\begin{multline}
\mathcal{A}_4^1 = \frac{1}{4(\ap)^2 (k_1 \cdot k_4) (k_2 \cdot k_3)} (u_1 u_4)(u_2 u_3) Y^{(cab)}Y^{(ba^{\parallel}c)} \\
G^{C_{bc}\bar{C}_{cb}} \int_0^{\infty} \d t \int_0^{it} \d q \langle V_0^{cb} (k,\theta^{\kappa},0) V_0^{bc} (-k,1-\theta^{\kappa},q) \rangle_{cc}
\end{multline}
plus permutations. The factors $Y^{(cab)}$ and $Y^{(ba^{\parallel}c)}$
are defined in the Appendix (\ref{YukDef}) - they are the
Yukawa couplings, derived entirely from tree-level
correlators in the desired basis. $G^{C_{bc}\bar{C}_{cb}}$ is the 
inverse of the K\"ahler metric $G_{C_{bc}\bar{C}_{cb}}$ for the chiral fields $C_{bc}$
in the chosen basis; note that we do not require its specific form, which was calculated in
\cite{Lust:scat,Bertolini:2005qh}. Thus we have reproduced the two Yukawa
vertices in the field-theory diagrams (figure \ref{fig:feynman}),
coupled with a propagator which contains the interesting information
about the running of the coupling. Note that if we had taken the limit
$z_1 \rightarrow z_2$, $z_4 \rightarrow z_3$ then we would factorise
onto a gauge propagator.

In the field theory the tree-level equivalent would have magnitude 
\begin{equation}
\mathcal{A}_4^0 = (u_1 u_4)(u_2 u_3)\frac{1}{2k_1 \cdot k_4}Y^{(cab)}Y^{(ba^{\parallel}c)}G^{C_{bc}\bar{C}_{cb}}
\end{equation}
while the one-loop diagram yields
\begin{equation}
\mathcal{A}_4^1 = (u_1 u_4)(u_2 u_3)\frac{1}{u^2}\Pi (u)Y^{(cab)}Y^{(ba^{\parallel}c)}(G^{C_{bc}\bar{C}_{cb}})^2
\end{equation}
where we put $u= 2k_1 \cdot k_4$ as the usual Mandelstam variable, and $\Pi (u)$ is the one-loop scalar propagator which we have yet to calculate 
(N.B. $\Pi(u) \sim G_{C_{bc}\bar{C}_{cb}}$). 
Thus if we want the renormalised Yukawa couplings in some basis 
(for example the basis where the fields are 
canonically normalised), we simply set $a^{\parallel}=a$ and write
\begin{equation}
(Y_R^{(cab)})^2 = (Y_{0}^{(cab)})^2(1 + \frac{1}{u}\Pi (u) G^{C_{bc}\bar{C}_{cb}} + \mathrm{permutations})
\label{YR}\end{equation}
where ``permutations'' accounts for the equivalent factors coming from the 
renormalisations of the fermion legs. 
Alternatively (and more precisely) 
since the superpotential receives no loop corrections, 
the above can be considered as a 
renormalisation of the K\"ahler potential
\begin{equation}
G_{R,C_{bc}\bar{C}_{cb}} = G_{C_{bc}\bar{C}_{cb}} (1 + \frac{1}{u}\Pi (u) G^{C_{bc}\bar{C}_{cb}})
\end{equation}

\subsection{The Scalar Propagator}

The object that remains to be calculated is of course the 
scalar propagator $\Pi (u)$ itself, which we can get from the following 
one-loop amplitude:
\begin{equation}
\Pi (k^2) = \int_0^{\infty} \d t \int_0^{it} \d q \langle V_0^{ab} (k,\theta^{\kappa},0) V_0^{ba} (-k,1-\theta^{\kappa},q) \rangle
\end{equation}
which represents wave-function renormalisation of the scalars in the
theory since as we have seen four-point chiral fermion amplitudes
will always factorise onto scalar two-point functions in the
field-theory limit. We fix both vertices to the same boundary along
the imaginary axis, and one of these we can choose to be at zero 
by conformal gauge-fixing. For the present, we shall specialise to the
following amplitude
\begin{equation}
\mathcal{A}_{2(a)}^{(ab)}= \int_0^{\infty} \d t \int_0^{it} \d q \langle V_0^{ab} (k,\theta^{\kappa},0) V_0^{ba} (-k,1-\theta^{\kappa},q) \rangle_{a^{\parallel}a}
\end{equation}
where the world sheet geometry is an annulus, and where one string end is
always fixed to brane $a^{\parallel}$, and the other is for some
portion of the cycle on brane $a$. In the latter region the 
propagating open string has untwisted boundary conditions 
so that in these diagrams the loop contains intermediate 
gauge bosons/gauginos. It is therefore these diagrams that will give 
Kaluza-Klein  (KK) mode contributions to the beta functions and power law 
running. The alternative (where the states {\em never} 
have untwisted boundary conditions) corresponds to only 
chiral matter fields in the loops, is harder to calculate and will 
be treated later and in Appendix C.

We have allowed the
fixed end of the string to be on a brane parallel to brane $a$ (rather than 
just $a$ itself), separated by a perpendicular 
distance $y^{\kappa}$ in each sub-torus. Diagrams where $y^{\kappa}\neq 0$ 
correspond to heavy stretched modes propagating in the loop and would 
in any case be extremely suppressed, but for the sake of generality 
we will retain $y^{\kappa}$.  

The diagram we are concentrating on here is present for any 
intersecting brane model, but in general $\Pi(k^2)$ receives 
contributions from other diagrams as well. For orientifolds, 
the full expression is 
\begin{equation}
\Pi (k^2) = \sum_{c} \mathcal{A}_{2(c)}^{(ab)} + \sum_{k,l} \left( M_{a,\Omega R \Theta^{k,l}a}^{(ab)} + M_{b,\Omega R \Theta^{k,l}b}^{(ab)} \right)
\label{FullPik2}\end{equation}
where $M^{(ab)}_{a,\Omega R \Theta^{k,l}a}$ is a M\"obius strip contribution, which we shall discuss later. Using the techniques described in Appendix \ref{LoopTech}, we find the amplitude
\begin{equation}
\mathcal{A}_{2(a)}^{(ab)} = \int_0^{\infty} \d t \int_0^{it} \d q A_{2(a)}^{(ab)} (q,t)
\end{equation}
where
\begin{multline}
A_{2(a)}^{(ab)} =2 \ap g_o^2 k^2 \tr (\gamma^a) \tr (\lambda^{(ab)} \lambda^{(ba)}) (8\pi^2 \alpha^{\prime}t)^{-2} \left[\frac{\theta_1 (q)}{\theta_1^{\prime}(0)} e^{-\frac{\pi}{t}(\Im (q))^2} \right]^{-2\alpha^{\prime} k^2}\\
\sum_{\nu=1}^4 N_{\nu} \delta_{\nu} \frac{\theta_{\nu} (0)}{\theta_1 (q)} \prod_{\kappa=1}^{3-d^{\prime}} \theta_1 (q)^{-\theta_{\kappa}} \theta_{\nu} (\theta^{\kappa}q) (L^{\kappa}T^{\kappa})^{-1/2} \\
\sum_{n_1^{\kappa}}\sum_{n_2^{\kappa}} \exp[-\frac{(D_A^{\kappa} (n_A^{\kappa}))^2 L^{\kappa}}{4\pi \alpha^{\prime} T^{\kappa}}] \exp[-\frac{(D_B^{\kappa} (n_B^{\kappa}))^2 T^{\kappa}}{4\pi \alpha^{\prime} L^{\kappa}}]
\label{Amplitude}\end{multline}
where the notation is as follows. The $\delta_{\nu} = \{1,-1,1,-1\}$ are the usual coefficients for
the spin-structure sum. The $\theta_\nu $ are the standard
Jacobi theta functions (see Appendix \ref{Thetas}) with modular
parameter  $it$ as on the worldsheet (so that we denote $\theta_{\nu} (z)
\equiv \theta_{\nu} (z, it)$ as usual). $D_A^{\kappa}$ and
$D_B^{\kappa}$ are the lengths of one-cycles, determined in the
Appendix to be
\begin{eqnarray}
D_A^{\kappa} (n_A^{\kappa}) &=& \frac{1}{\sqrt{2}}n_A^{\kappa} L_a^{\kappa} \nonumber \\
D_B^{\kappa} (n_B^{\kappa}) &=& n_B^{\kappa} \sqrt{2}\frac{4\pi^2 T_2^{\kappa}}{L_a^{\kappa}} + y^{\kappa}
\end{eqnarray}
where $L_a^{\kappa}$ is the wrapping cycle length of brane $a$ in
sub-torus $\kappa$, and $T_2^{\kappa}$ is the K\"ahler modulus for the
sub-torus, given by $R_1^{\kappa}R_2^{\kappa}\sin \alpha^{\kappa}$,
$R_1^{\kappa}$ and $R_2^{\kappa}$ are the radii of the torus,
$\alpha^{\kappa}$ is the tilting parameter, and generally
\[L_a^{\kappa} = 2\pi \sqrt{(n_a^{\kappa}R_1^{\kappa})^2 +
(m_a^{\kappa}R_2^{\kappa})^2 +
2n_a^{\kappa}m_a^{\kappa}R_1^{\kappa}R_2^{\kappa}\cos
\alpha^{\kappa}}. \]
$N_\nu$ are normalisation factors that we will determine in the next subsection.
Finally the classical (instanton) sum depends
on two functions $L^\kappa $ and $T^\kappa $ 
\begin{eqnarray}
L^{\kappa}(q,\theta^{\kappa}) &=& \int_{-\frac{1}{2}}^{\frac{1}{2}} \d z \omega_1(z)\nonumber \\
T^{\kappa}(q,\theta^{\kappa}) &=& \int_{-\frac{1}{2} + it}^{-\frac{1}{2}} \d z \omega_1 (z) 
\end{eqnarray}
where
\begin{equation}
\omega_1 (z) = \left( \frac{\theta_1(z)}{\theta_1(z-q)} \right)^{\theta^{\kappa}-1} \frac{\theta_1 (z-\theta^{\kappa}q)}{\theta_1 (z-q)}
\end{equation}
and where the contour for $L^{\kappa}$ is understood to pass under the
branch cut between $0$ and $q$. Note that this contrasts with the four
independent functions that we might expect from the equivalent closed
string orbifold calculation.

\subsection{Factorisation on Partition Function}

The normalisation factors for the amplitude $N_\nu$  must still be determined 
 by factorising on the partition function (i.e. bringing the remaining 
two vertices together to eliminate the branch cuts entirely) 
and using the OPE coefficients of the various CFTs.  We should obtain
\begin{equation}
A_{2(a)}^{(ab)}  \sim q^{-1-\sum_{\kappa}\theta^{\kappa}}2\alpha^{\prime} k^2 g_o^2  \tr (\gamma^a) \tr (\lambda^{(ab)} \lambda^{(ba)}) Z_{aa}(it) \prod_{\kappa=1}^3 C^{(aba)0}_{\theta^{\kappa}, 1-\theta^{\kappa}}
\end{equation}
for $\sum_{\kappa}\theta^{\kappa} \le 1$, where $g_o$ is the open string coupling, 
$C^{(aba)0}_{\theta^{\kappa}, 1-\theta^{\kappa}}$ is the OPE coefficient of the boundary-changing 
operators determined in Appendix \ref{CFT} and 
$Z_{aa}(it)$ is the partition function for brane $a$. It is given by
\begin{equation}
Z_{aa}(it) = (8\pi^2 \alpha^{\prime}t)^{-2} \eta (it)^{-12} \sum_{\nu=1}^{4} \delta_{\nu} \theta_{\nu}^4 (0) \prod_{\kappa=1}^3 Z_a^{\kappa} 
\end{equation}
where
\begin{equation}
Z_{a}^{\kappa} (t,T^{\kappa}_2,L_a^{\kappa}) = \sum_{r^{\kappa},s^{\kappa}} \exp \bigg[ \frac{-8\pi^3 \alpha^{\prime} t}{(L_a^{\kappa})^2} |r^{\kappa} + \frac{iT_2^{\kappa} s^{\kappa}}{\alpha^{\prime}}|^2 \bigg]
\end{equation}
is the bosonic sum over winding and kaluza-klein modes. The factorisation occurs for $q \rightarrow 0$, when $L^{\kappa} \rightarrow 1$ and $T^{\kappa} \rightarrow it$, giving
\begin{equation}
N_{\nu} = \frac{e^{\Phi}}{\alpha^{\prime}g_o^2} (2\pi)^3 (\sqrt2)^{-3+d^{\prime}} \left(\prod_{\iota=1}^{d^{\prime}} \theta_{\nu} (0)Z_a^{\iota} \right)\eta^{-6} (it) G_{C_{ab}\bar{C}_{ba}}
\end{equation}

\section{Divergences}

With our normalised amplitude, we are finally able to probe its
behaviour. The important limits are $q \rightarrow 0$ and $q
\rightarrow it$ where there are poles; in the former case the pole is
cancelled because of the underlying $N=4$ structure of the gauge sector, 
but in the latter it is not, and it dominates the
amplitude. We should comment here about a subtlety with these
calculations which does not apply for many other string amplitudes:
due to the branch cuts on the worldsheet, the amplitude is not
periodic on $q \rightarrow q + it$. This causes the usual prescription
of averaging over the positions of fixed vertices to give a symmetric
expression to break down (this was an ambiguity in \cite{Ben}): the
gauge-fixing procedure asserts that one vertex \emph{must} be fixed,
which, to remain invariant as $t$ changes, is placed at zero. The
non-periodic nature of the amplitude is entirely expected from the
boundary conformal field theory perspective: due to the existence of a
non-trivial homological cycle on the worldsheet, we have two OPEs for
the boundary changing operators, depending on how (i.e. whether) we
combine the operators to eliminate one boundary. We have already used
the expected behaviour in the limit $q \rightarrow 0$ to normalise the 
amplitude, but $q \rightarrow it$ yields new information, namely 
the partition function for string stretched between different branes:
in this limit we obtain

\begin{equation}
A_{2(a)}^{(ab)}  \sim (it-q)^{-1-\sum_{\kappa}\theta^{\kappa}-2\alpha^{\prime}k^2} 2 \ap
k^2 g_o^2  \tr (\gamma^a) \tr (\lambda^{(ab)} \lambda^{(ba)}) Z_{ab}(it) \prod_{\kappa=1}^3 
C^{(bab)0}_{\theta^{\kappa}, 1-\theta^{\kappa}}\, ,
\end{equation}
where $Z_{ab}(it)$ is the partition function for string stretched between branes $a$ and $b$, given by
\begin{equation}
Z_{ab}(it) = (8\pi^2 \alpha^{\prime}t)^{-2} \bigg(\prod_{\iota=1}^{d^{\prime}} Z_a^{\iota} \bigg) \sum_{\nu} \delta_{\nu} \left(\frac{\theta_{\nu} (0)}{\eta^3(it)}\right)^{1+d^{\prime}} \prod_{\kappa=1}^{3-d^{\prime}}I_{ab}^{\kappa} \frac{\theta_{\nu} (i\theta^{\kappa}t)}{\theta_{1} (i \theta^{\kappa} t)}
\label{Zab}\end{equation}

For supersymmetric configurations, however, this partition function is zero, and hence we were required to calculate our correlator to find the behaviour in this limit. It is straightforward to show that our calculation gives this behaviour: as $q \rightarrow it$, the functions $L^{\kappa}$ diverge logarithmically, and so we perform a poisson resummation over $n_B^{\kappa}$ - which reduces the instanton sums to unity. Simple complex analysis gives
\begin{equation}
T^{\kappa} = \int_{it}^{q} \d z \omega_1 (z) \rightarrow B(\theta^{\kappa},1-\theta^{\kappa})\exp [-\pi\theta^{\kappa}t] \frac{\theta_1 (\theta^{\kappa} it)}{\theta_1^{\prime}(0)}
\end{equation}
and we also require the identity \cite{Lust:Gauge}:
\begin{equation}
\sin \pi \theta^{\kappa} = \frac{4\pi^2 T_2^{\kappa} I_{ab}^{\kappa}}{L_a^{\kappa}L_b^{\kappa}}
\end{equation}
so that, for the $N=1$ supersymmetric choice of angles we obtain (after Riemann summation):
\begin{equation}
A_{2(a)}^{(ab)} (q,t) \sim (it-q)^{-1-2\alpha^{\prime}k^2} 2
k^2 e^{\Phi} \tr (\gamma^a) \tr (\lambda^{(ab)} \lambda^{(ba)}) 
\frac{I_{ab}}{L_b}2\pi (2\pi \sqrt{\alpha^{\prime}})^3 
G_{C_{ab}\bar{C}_{ba}}(8\pi^2 \alpha^{\prime}t)^{-2}
\label{DivPole}\end{equation}
The above is clearly singular in the limit $k^2 \rightarrow 0$, where the integral over $q$ is dominated by the behaviour at $q=it$, since at $q=0$ the effective $N=4$ SUSY of the gauge bosons cancels the pole. Using the usual prescription for these integrals we obtain the exact result for the amplitude, and find it is divergent:

\begin{equation}
\int_0^{it} \d q A_{2(a)}^{(ab)} (q,t)= \frac{e^{\Phi}}{4(\alpha^{\prime})^{3/2}}  G_{C_{ab}\bar{C}_{ba}}\tr(\lambda^{(ab)} \lambda^{(ba)})\frac{N_a I_{ab}}{L_b}\int_0^{\infty} \d t t^{-2}.
\label{Divergence}\end{equation}

This divergence is effectively due to $RR$-charge exchange between the
branes, and so we look to contributions from other diagrams (as given
in (\ref{FullPik2})) to cancel it. These consist of other annulus
diagrams where one end resides on a brane other than ``a'' or ``b'',
and M\"obius strip diagrams. 

Fortunately we do not need to calculate
the amplitude of these additional contributions in their entirety; indeed
we can obtain all that we need from
knowledge of the partition functions and the properties of our
boundary changing operators, along with a straightforward conjecture about
the behaviour of the amplitude. 
For an annulus diagram where one string end remains on a brane ``c'',
while the other end contains the vertex operators and thus is attached
to brane ``a'' or ``b'', we obtain two contributions: one from each
limit. From the OPEs of the boundary-changing operators, we expect to
obtain
\begin{equation}
A^{(ab)}_{2(c)} (x,t)= 2 \ap g_{0}^2 k^2 \tr (\gamma^{c}) \tr (\lambda^{(ab)} \lambda^{(ba)})(x)^{-2-2\alpha^{\prime}k^2} \bigg( C^{(aba)} Z_{ac} + C^{(bab)} Z_{bc} \bigg)
\end{equation}
where $x$ now denotes $q$ or $it-q$ in the appropriate limits, and
$C^{(aba)}$ is understood to be the product of the OPE coefficients
for each dimension. Considering the partition functions (\ref{Zab})
and the behaviour of the expression (\ref{DivPole}), then we propose
that the effect of the division by zero in the above is to cancel one
factor of $x$ with a factor of $\theta_1(0)$. Hence, if we write
equation (\ref{Divergence}) as $\mathcal{A}_0 \frac{N_a I_{ab}}{L_b}$,
we thus obtain for the divergence,

\begin{equation}
\mathcal{A}_{2(c)}^{(ab)} = \mathcal{A}_0 \left ( \frac{N_c I_{ac}}{L_a} + \frac{N_c I_{bc}}{L_b} \right)
\end{equation}
Note that $c$ is allowed to be any brane in the theory, including images under reflection and orbifold elements. 

We must now also consider the contribution from M\"obius strip diagrams, which, with the same conjectured behaviour as above, give us
\begin{multline}
M_{a,\Omega R \Theta^{k,l}}^{(ab)} = 2\ap g_{0}^2 k^2 tr(\lambda_1 \lambda_1^{\dagger} (\gamma^{\Omega R \Theta^{k,l}a}_{\Omega R \Theta^{k,l}})^* \gamma^{a}_{\Omega R \Theta^{k,l}} ) q^{-2-2\alpha^{\prime}} \\
\bigg( C^{(aba)} M_{a,\Omega R \hat{\Theta}^{k,l}a} + C^{(bab)} M_{b,\Omega R \hat{\Theta}^{k,l}b} \bigg)
\end{multline} 
where $M_{a,\Omega R \Theta^{k,l}a}$ denotes the M\"obius diagram between brane $a$ and its image under the orientifold group $\Omega R \hat{\Theta}^{k,l}a$, supplemented by a twist insertion $\hat{\Theta}^{k,l}$ to give a twist-invariant amplitude, given by \cite{Lust:Gauge}
\begin{multline}
M_{a,\Omega R \Theta^{k,l}a} = -(8\pi^2 \alpha^{\prime}t)^{-2} \left( \prod_{\iota=1}^{d^{\parallel}} n^{\iota}_{O6_{k,l}} L_i (t,T^{\iota}_2,L^{\iota}_{O6_{k,l}})\right) \\
\sum_{\nu} \delta_{\nu} \left(\frac{\theta_{\nu} (0,it+\frac{1}{2})}{\eta^3(it+\frac{1}{2})}\right)^{1+d^{\parallel}} \prod_{\kappa=1}^{3-d^{\parallel}} 2^{\delta^{\kappa}} I^{\kappa}_{a,O6_{k,l}}  \frac{\theta_{\nu} (2\theta^{\kappa}it,it+\frac{1}{2})}{\theta_1(2\theta^{\kappa}it,it+\frac{1}{2})},
\end{multline} 
where $\delta^{\kappa}$ \cite{Blumenhagen:1999ev} is zero for $a$
orthogonal to the $O6_{k,l}$-plane in sub-torus $\kappa$, and 1
otherwise; $d^{\parallel}$ is the number of sub-tori in which brane
$a$ lies on the $O6_{k,l}$-plane, and $n^{\iota}_{O6_{k,l}}$ is the
number of times the plane wraps the torus with cycle length
$L^{\iota}_{O6_{k,l}}$. Here $\theta^{\kappa}$ is the angle between
brane $a$ and the $O6_{k,l}$-plane. We also define
$I^{\kappa}_{a,O6_{k,l}}$ to be the total number of intersections of
brane $a$ with the $O6$-planes of the \emph{class} $[O6_{k,l}]$ in
sub-torus $\kappa$. For $d^{\parallel}$ non-zero we have zero modes:
\begin{equation}
L_i (t,T^{\iota}_2,L^{\iota}_{O6_{k,l}}) = \sum_{r^{\iota},s^{\iota}} \exp-\frac{8\pi^3 \alpha^{\prime} t}{T_2^{\iota} L^{\iota}_{O6_{k,l}}} | r^{\iota} + \frac{i2^{\mu}T^{\iota}_2 s^{\iota}}{\alpha^{\prime}} |^2
\end{equation}
where $\mu=0$ for untilted tori, and $1$ for tilted tori. To 
obtain the divergence, the same procedure as before can be applied 
once we have taken into account the relative scaling of the 
modular parameter between the M\"obius and annulus
diagrams. Since the divergence occurs in the ultraviolet and is due
to closed string exchange, to do this we transform to the
closed string channel: we simply replace $t$ by
$1/l$ for the annulus, and $1/(4l)$ for the M\"obius strip. This
results in an extra factor of $4$ preceding the M\"obius strip
divergences relative to those of the annulus diagrams, which are due to the
charges of the $O6$-planes being $4$ times those of the $D6$-branes. We
thus obtain the total divergence
\begin{multline}
\mathcal{A}_{2}^{1}=\frac{e^{\Phi}}{4(\alpha^{\prime})^{3/2}}  G_{C_{ab}\bar{C}_{ba}}\tr(\lambda^{(ab)} \lambda^{(ba)})\int_0^{\infty} \d l \bigg\{\frac{1}{L_a} \left ( \sum_{c,c^{\prime}} N_c I_{ac} + 4\sum_{k,l} \rho_{\Omega R \hat{\Theta}^{k,l}} I_{a,O6_{k,l}} \right) \\
+ \frac{1}{L_b} \left ( \sum_{c,c^{\prime}} N_c I_{bc} - 4\sum_{k,l} \rho_{\Omega R \hat{\Theta}^{k,l}} I_{b,O6_{k,l}} \right) \bigg\}
\end{multline}

Note that the total divergence is of the same form as that found in gauge
coupling renormalisation. We recognise the terms in brackets as the
standard expression for cancellation of anomalies, derived from the
$RR$-tapole cancellation conditions \cite{Cvetic:Models}:
\begin{equation}
\label{noname}
[\Pi_a] \cdot \left( \sum_{c,c^{\prime}} N_{c} [\Pi_c] -4 \sum_{k,l} [\Pi_{O6^{k,l}}] \right) = 0
\end{equation}
where $[\Pi_a]$ is the homology cycle of $a$ etc; note that the phases $\rho_{\Omega R \hat{\Theta}^{k,l}}$ 
will be the same as the sign of the homology cycles of the orientifold planes.
Hence, we have shown that cancellation of $RR$-charges implies that in the limit 
that $k^2 \rightarrow 0$, the total two-point amplitude is zero.

\section{Running Yukawas up to the String Scale}

Having demonstrated the
mechanism for cancellation of divergences in the two point function, 
we may now analyse its
behaviour and expect to obtain finite results. As mentioned earlier,
by examining the amplitude at small, rather than zero, $k^2$, we
obtain the running of the coupling as appearing in four-point and
higher diagrams where all Mandelstam variables are not necessarily
zero. Unfortunately, we are now faced with two problems: we have only
calculated the whole amplitude for one contribution; and an exact
integration of the whole amplitude is not possible, due to the complexity
of the expressions and lack of poles. However, this does not prevent
us from extracting the field-theory behaviour and even the running
near the string scale, but necessarily involves some approximations.

Focusing on our amplitude (\ref{Amplitude}), which in the field theory
limit comprises the scalar propagator with a self-coupling loop and a
gauge-coupling loop, we wish to extract the dependence of the entire
amplitude upon the momentum. Schematically, according to equation
(\ref{YR}), we expect to obtain for Yukawa coupling $Y$, as $k^2
\rightarrow 0$,

\begin{equation}
Y_R - Y_0 \sim A + B + \frac{g^2\beta}{8\pi^2} \ln k^2 + \Delta + O(k^2)
\end{equation}
where $A$ represents the divergent term, the third term is the
standard beta-function running, and the fourth term comprises all of
the threshold corrections. This is the most interesting term: as
discussed in \cite{Ben} it contains power-law running terms, but with
our complete expressions here we are able to see how the
running is softened at the string scale. The term denoted $B$ is a possible 
finite piece that is zero in supersymmetric configurations but 
that might appear in non-supersymmetric ones: in the field
theory it would be proportional to the cutoff squared while in the
(non-supersymmetric) string theory it would be finite. This term
would give rise to the hierarchy problem. Had we found such a term
in a supersymmetric model it would have been
inconsistent with our expectations about the tree-level K\"ahler
potential - in principle, it could only appear if there were not total
cancellation among the divergent contributions.

The power-law running in the present case corresponds to 
both fermions and Higgs fields being localised at
intersections, but gauge fields having
KK modes for the three extra dimensions on the 
wrapped D6 branes \cite{DDG}. For completeness we write the 
expression for KK modes on brane $a$ 
with three different KK thresholds $\mu_{0,1,2}$ 
(one for each complex dimension - note that in principle we have 
different values for each brane $a$):
\begin{equation}
\Delta = \frac{g_a^2}{8\pi^2} \left( (\beta - \hat{\beta}) \ln \frac{\Lambda}{\mu_0} + \hat{\beta} \sum_{\delta=1}^3 \frac{X_{\delta}}{\delta} \left[ \left(\frac{\mu_{\delta}}{\mu_{\delta-1}}\right)^{\delta} -1 \right] \prod_{i<\delta} \left( \frac{\mu_i}{\mu_{i-1}} \right)^{i}
\right) + \Delta_S
\end{equation} 
where $\{ \mu_{\delta} \} = \{ (L_a^{\delta+1})^{-1}, \Lambda |\
\mu_0 < \mu_1 < \mu_2 < \Lambda\}$, and $\Lambda$ is the string cutoff
which should be ${\cal O}(1)$ for our calculation. $\{X_{\delta}\} =
\{2,\pi,4\pi/3\}$ is the correction factor for the sum, and $\Delta_S$ is the
string-level correction. $\beta$ and $\hat{\beta}$ are the
beta-coefficients for the standard logarithmic running and power-law
running respectively. Note that the above is found from an integral
over the schwinger parameter $t^{\prime}$ where the integrand varies 
as $(t^{\prime})^{-\delta/2-1}$ in each 
region; $t^{\prime}$ is equivalent to the string modular parameter $t$, modulo
a (dimensionful) proportionality constant.

To extract the above behaviour, while eliminating the divergent term,
without calculating all the additional diagrams, we could impose a
cutoff in our $q$ integral (as in \cite{Ben}); however the physical meaning of 
such a cut-off is obscure in the present case, so we shall not do that here. As in
\cite{Ben}, we shall make the assumption that the classical sums are
well approximated by those of the partition function for the gauge
boson. However, we then subtract a term from the factors preceding the
classical sum, which reproduces the pole term with no subleading
behaviour in $k^2$; since the region in $q$ where the Poisson
resummation is required is very small, this is a good way to regulate
the pole that we have in the integrand when we set $k^2$ to zero. To
extract the $\Delta$ terms, we can set $k^2$ to zero in the integrand:
this is valid except for the logarithmic running down to zero energy
(i.e. large t), where the $k^2$ factors in the $\left[\frac{\theta_1
(q)}{\theta_1^{\prime}(0)} e^{-\frac{\pi}{t}(\Im (q))^2}
\right]^{-2\alpha^{\prime} k^2}$ term regulate the remaining $t^{-1}$
behaviour of the integrand when the $t$ integral is performed. This
behaviour is then modified by powers of
$(8\pi^2\alpha^{\prime}t/(L_a^{\kappa})^2)^{1/2}$ multiplying the
classical sums after each Poisson-resummation, as $t$ crosses each cutoff
threshold, yielding power-law running as expected; this was obtained
in \cite{Ben}, and so we shall not reproduce it here. 

\begin{figure}\begin{center}
\epsfig{file=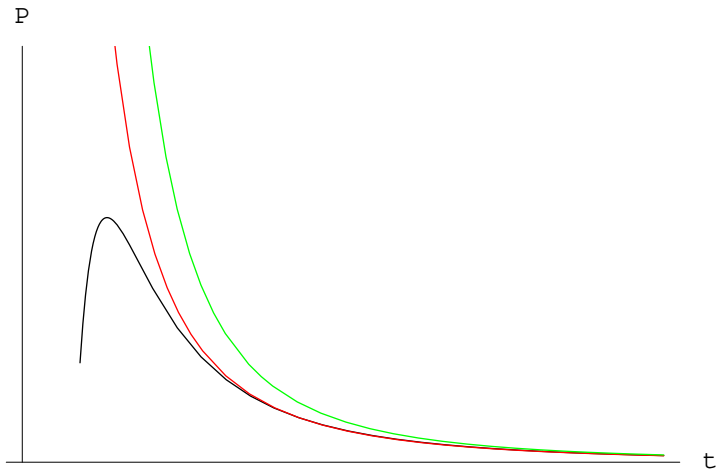}
\caption{A linear plot of the ``running'' Yukawa coupling in modular
parameter $t$, lower graph. The peak is very close to
$t=1$, i.e. the string scale. The top graph is the standard power-law
behaviour continued. The middle graph is the field theory 
approximation using string improved propagators as defined in the
text.}\label{running}
\end{center}
\end{figure}

For $\Lambda^{-2} > t \sim 1$, we have an intermediate stage where the
KK modes are all excited, but we have not yet excited the
winding modes, represented by the sums over $n_B^{\kappa}$. 
As a first approximation, the integral, with our regulator term, is
\begin{multline}\label{graphcalc}
\mathcal{A}_{2(a)}^{(ab)} \sim e^{\Phi} \frac{k^2}{8\pi (\alpha^{\prime})^2
\sqrt{2}} G_{C_{ab}\bar{C}_{ba}} \int_0^{\infty} \d t \int_0^{t} \d
\lambda t^{-7/2} \eta^{-6} (it) \\ \left( \frac{\theta_1 (\theta^1
i\lambda) \theta_1 (\theta^2 i\lambda)\theta_1 (\theta^3
i\lambda)}{\theta_1 (i\lambda)} + \frac{\theta_1 (\theta^1 it)
\theta_1 (\theta^2 it)\theta_1 (\theta^3
it)}{\theta_1^{\prime}(it)}\left( \frac{1}{(t-\lambda)} -
\frac{1}{t}\ln t \right) \right)
\end{multline}   
where we have used $L^{\kappa} (i\lambda,it) T^{\kappa} (i\lambda,it) \approx it$.
If we had not set $k^2$ to zero in the integrand, the second term in brackets would be \[
i^{-2\ap k^2}\left((t-\lambda)^{-1-2\ap k^2}+\frac{t^{-2\alpha^{\prime}k^2}-1}{2\alpha^{\prime}k^2t}\right).
\] The above can then be integrated numerically to give 
\begin{equation}
\mathcal{A}_{2(a)}^{(ab)} \sim e^{\Phi} \frac{k^2}{8\pi (\alpha^{\prime})^2 \sqrt{2}} G_{C_{ab}\bar{C}_{ba}} \int_0^{\infty} \d t P(t)
\end{equation}
where $P(t)$ is plotted in figure \ref{running}. It indicates how the
threshold corrections are changing with energy scale probed.  
The figure clearly demonstrates the softening of the running near
the string scale. 
Alas, we also find that the amplitude still diverges (negatively) 
as $t\rightarrow 0$, and so we conclude that the 
subtraction of the leading poles, rather than fully 
including the remaining pieces (i.e. the other annulus diagrams 
and the M\"obius strip diagrams), was not enough to render a 
finite result. However we believe that the condition 
in eq.(\ref{noname}) ensures cancellation of the
remaining divergences as well.

Despite this, the procedure of naive pole cancellation still enhances our
understanding of the running up to the string scale,
because the divergent pieces (depending as they do on much heavier modes) 
rapidly die away as $t$ increases. This allows us to 
focus on the behaviour of the KK contribution.
The quenching of UV divergent KK contributions is well
documented at tree-level but has not been discussed in detail 
at one-loop. At tree-level the nett effect of the string theory is to 
introduce a physical Gaussian cut-off in the infinite sum over 
modes. For example a tree level $s$-channel 
exchange of gauge fields is typically proportional to 
\cite{Antoniadis:2000jv,Abel:2003fk}
\begin{equation}
\sum_{\underline{n}}
\frac{\prod_\kappa e^{-\beta^\kappa M_{n^\kappa}^2 \alpha'}}
{s-M_{\underline{n}}^2},
\end{equation}
where $\beta^\kappa = 
( 2\psi(1) - \psi(\theta^\kappa)-\psi(1-\theta^\kappa) ) $ and where 
$M_{\underline{n}}^2 = \sum_\kappa M^2_{n^\kappa} $ is the 
mass-squared of the KK mode. Note that the above is similar to results in \cite{Dudas:1999gz,Cullen:2000ef}.
The physical interpretation of this expression is that the 
D-branes have finite thickness of order $\sqrt{\alpha'}$. Consequently 
they are unable to excite modes with a shorter Compton wavelength than 
this which translates into a cut-off on modes whose mass is greater than 
the string mass. This suggests the possibility of ``string improved propagators'' 
for the field theory; for example scalars of mass $m$ would have 
propagators of the form
\begin{equation}
\Delta(k)=\frac{\prod_\kappa e^{-\beta^\kappa (m^\kappa)^2 \alpha'}}{k^2-m^2},
\end{equation}
where $m^2 = \sum_\kappa (m^\kappa)^2 $, and we are neglecting gauge invariance. 
To test this expression we can follow
through the field theory analysis in ref.\cite{DDG} (which also neglects
gauge invariance): the power law running is modified:
\begin{equation}
\frac{1}{t} t^{-\frac{\delta}{2}} \rightarrow 
\frac{1}{t} \prod_\kappa (t+\frac{\beta^\kappa}{\pi} )^{-\frac{1}{2}}. 
\end{equation}
For comparison this curve is also included in figure \ref{running}. Clearly 
it provides a better approximation to the string theory up to near the 
string scale, justifying our propagators. (Of course the usual logarithmic 
divergence in the field theory remains once the KK modes are all quenched.)
Also interesting is that this behaviour appears in our first approximation; 
we expect it to appear due to $T^{\kappa} (i\lambda,it)$ and
$L^{\kappa} (i\lambda,it)$ differing from the values we have assumed, where we would
also have to take into account the poisson resummation required near 
$\lambda \rightarrow t$. We could thus attempt an improved approximation
by modifying these quantities: however, the integrals rapidly become
compuationally intensive, and we shall not pursue this further here.

\section{Further Amplitudes}

Having discussed the one-loop scalar propagator and its derivation from
$N$-point correlators of boundary-changing operators, we may wish
to consider calculating the full $N$-point amplitudes themselves. 
The general procedure was outlined in \cite{Ben} where 
three-point functions were considered. The complete procedure 
for general $N$-point functions is presented in
Appendix \ref{LoopTech}, the main new result being equation (\ref{M}).

Furthermore, we may also wish to calculate diagrams in which there are
only chiral matter fields in the loop; although we were able to 
use the OPEs to obtain the necessary information pertaining 
to the divergences, we would
need these diagrams to study the full
running of the scalar propagator, for example. We have constructed a
new procedure for calculating them, and describe it in full
in Appendix \ref{AdvancedLoopTech}; the crucial step is to modify the
basis of cut differentials.

In calculating the contribution to the scalar propagator from diagrams
with no internal gauge bosons, we find some intriguing behaviour. In
the field theory, the diagrams which correspond to these (other than
the tadpole) involve two Yukawa vertices, and hence we should expect
the result to be proportional to products of Yukawa couplings; 
we expect to find the angular
factors and the instanton sum. We already see from the CFT analysis
that the divergences are not proportional to these which is perhaps 
not surprising. However when we calculate the finite piece of diagrams, 
although we do obtain the expected instanton
terms in the classical action it is not immediately obvious that we
obtain the Yukawa couplings here either. Yukawas
attached to external legs will always appear by 
factorisation thanks to the 
OPE, but this is not the case for internal Yukawas. 
Hence, it seems likely that the
field theory behaviour of composition of amplitudes is only
approximately exhibited at low energy, and near the string scale
this breaks down. This will be a source of flavour changing 
in models that at tree-level have flavour diagonal couplings.
Unfortunately it cannot solve the ``rank 1'' problem of 
the simplest constructions however, because canonically normalizing 
the fields (i.e. making the K\"ahler metric flavour diagonal) 
does not change the rank of the Yukawa couplings.
We leave the full analysis of this to future work.

\section{Conclusions}

We have presented the procedure for calculating $N$-point diagrams in
intersecting brane models at one-loop. 
We have shown that the cancellation of 
leading divergences in the scalar
propagator for self and gauge couplings for supersymmetric 
configurations is guaranteed by $RR$-tadpole 
cancellation.
The one-loop correction to the propagator is 
consistent with a canonical form of the 
K\"ahler potential in the field theory, or one of the no-scale variety, 
where there are no divergences in the field theory; had there 
been a constant term
remaining, this would have corresponded to a divergence in the field-theory
proportional to the UV cut-off squared, which would have been consistent with
alternative forms of the K\"ahler potential. 
However, other than the expected corrections from
power-law running, we cannot make specific assertions for corrections
to the K\"ahler potential from the string theory.

When we investigated the energy dependence of the scalar propagator
(in the off-shell extension, i.e. as relevant for the four-point and
higher diagrams) we found that there still remained divergences, which
can only be cancelled by a full calculation of all the diagrams in the
theory. However information could be obtained about the intermediate 
energy regime where KK modes are active and affecting the running. We 
find the tree level behaviour whereby the string theory quenches
the KK modes remains in the one-loop diagrams, and we proposed 
a string improved propagator that can take account of this in the
field theory.

We also developed the new technology necessary for calculating
annulus diagrams without internal gauge bosons, and mention some
interesting new features. At present, however, the technology for
calculating the M\"obius strip diagrams does not exist, and this is
left for future work.

\section*{Acknowledgements}
We would like to thank Daniel Roggenkamp and Mukund Rangamani for some valuable discussions. M. D. G. is supported by a PPARC  studentship.

\appendix

\section{Boundary-Changing Operators}\label{CFT}

The technically most significant part of any calculation involving chiral fields stretched between branes is the manipulation of the boundary-changing operators. They are primary fields on the worldsheet CFT which represent the bosonic vacuum state in the construction of vertex operators, so we require one for every stretched field. In factorisable setups such as we consider in this paper we require one for every compact dimension where there is a non-trivial intersection - so we will require three for each vertex operator in N=1 supersymetric models. 

In previous papers (e.g. \cite{Cvetic:Yukawa,Owen}) the boundary-changing operator for an angle $\theta$ has been denoted $\sigma_{\theta}$, in analogy with closed-string twist-field calculations. It has conformal weight $h_{\theta} = \frac{1}{2} \theta (1-\theta)$, and we can thus write the OPE of two such fields as
\begin{equation}
\sigma_{\nu} (z_1) \sigma_{\lambda} (z_2) \sim \sum_{k} C_{\nu \lambda}^{k} \sigma_{k}(z_2) (z_2 - z_1)^{h_k - h_{\nu} - h_{\lambda}}
\end{equation}
Moreover, the ``twists'' are additive, so $k=\nu + \lambda$ for $\nu + \lambda < 1$, or $k=\nu + \lambda -1$ for $\nu + \lambda > 1$. However, they also carry on the worldsheet labels according to the boundaries that they connect; for a change from brane $a$ to $b$ we should write $\sigma_{\theta}^{(ab)}$. We then have non-zero OPEs only when the operators share a boundary, so that we modify the above to 
\begin{equation}
\sigma_{\nu}^{(ab)} (z_1) \sigma_{\lambda}^{(cd)} (z_2) \sim \delta^{bc} C_{\nu \lambda}^{(abd)\nu + \lambda} \sigma_{\nu + \lambda}^{(ad)}(z_2) (z_2 - z_1)^{h_{\nu + \lambda} - h_{\nu} - h_{\lambda}}
\end{equation}

We also find that new fields appear, representing strings stretched between parallel branes; the OPE coefficients and weights can all be found by analysing tree-level correlators. In order to normalise the fields and thus the OPEs we must compare the string diagrams to the low-energy field theory, specifically the Dirac-Born-Infeld action for the $IIA$ theory:

\begin{equation}
S_6 = -T_6 \int \d^{7} \xi e^{-\Phi} tr \sqrt{-\det(G+B+2\pi\alpha^{\prime}F)}
\end{equation}
where $G$ and $B$ are pull-backs of the metric and antisymmetric tensor to the brane world-volume, $F$ is the gauge field strength, and $e^{\Phi}$ is the closed string coupling. If we consider gauge excitations only on the non-compact dimensions, the effective field theory has gauge kinetic function \cite{Lust:scat,Cvetic:Yukawa}:
\begin{equation}
g^{-2}_{D6} = e^{-\Phi} \frac{1}{2\pi} \prod_{\kappa=1}^{3} \frac{L_a^{\kappa}}{2\pi\sqrt{\alpha^{\prime}}} 
\end{equation}
where we have the volume of brane $a$ equal to $L_a = \prod_{\kappa=1}^3 L_a^{\kappa}$. If we now consider two, three and four-point gauge-boson amplitudes (for non-abelian gauge group) in the 4d effective theory, they must all have the same coupling; this is only possible if the gauge kinetic function is associated to the normalisation of the disc diagram, and not the fields, implying
\begin{equation}
(\alpha^{\prime})^{-1} g^{-2}_o \bra 1 \ket_a = g^{-2}_{a}
\end{equation}
where $g_o$ is the open string coupling. Clearly the gauge bosons must be canonically normalised by a factor of $g_o^{-1}$ to match the field theory, where the coupling is also absorbed into the fields. Note that the above can also be obtained by a boundary state analysis without recourse to the low energy supergravity \cite{Gaberdiel,Dorey}.

 We now consider the two-point function for boundary-changing operators; we require
\begin{equation}
\bra \sigma_{\theta}^{ab} \sigma_{1-\theta}^{ba} \ket = G_{C_{\theta}\bar{C}_{\theta}} = C^{(aba)0}_{\theta, 1-\theta} \bra 1 \ket_a 
\end{equation}
where $G_{C_{\theta}\bar{C_{\theta}}}$ is the K\"ahler metric of the chiral multiplets \cite{Lust:scat}. This allows us to determine the OPE coefficients; however, for convenience we shall combine them (for now) with the string coupling since that is how they shall always appear:
\begin{equation}
(g_o)^2 \prod_{\kappa=1}^{3} C^{(aba)0}_{\theta^{\kappa}, 1-\theta^{\kappa}} = (\alpha^{\prime})^{-1} g^2_a G_{C_{ab}\bar{C}_{ba}}
\end{equation}
To obtain further correlators we must now consider four-point tree diagrams. First we consider $\mathcal{A}_{tree} \equiv g_o^4 \bra \sigma_{\nu}^{ab} (z_1) \sigma_{1-\nu}^{ba^{\parallel}} (z_2) \sigma_{1-\lambda}^{a^{\parallel}c} (z_3) \sigma_{\lambda}^{ca}(z_4)\ket_{D_2}$, which was calculated for a particular limit in \cite{Cvetic:Yukawa,Lust:scat}, but for the general case in \cite{Owen} (note that the product over dimensions is implied). The result was 
\begin{multline}
\mathcal{A}_{tree}= C (\alpha^{\prime})^{-1} g_o^2 \prod_{\kappa=1}^3 z_{12}^{-2h_{\nu^{\kappa}}} z_{34}^{-2h_{\lambda^{\kappa}}} I^{\kappa}(x)^{-1/2} \left(1-x \right)^{ - \nu^{\kappa} \lambda^{\kappa} } \\
\sum_{n_1,n_2} \exp \left[\frac{-\sin \pi\nu}{4\pi\alpha^{\prime}} \frac{(v_{21}\tau - v_{32})^2 + (v_{21}\tau^{\prime} + v_{32})^2}{\tau + \tau^{\prime}} \right]
\end{multline}
where $v_{21}(n_1) = L_{0}^{21,\kappa} + n_1 L_b^{\kappa}$ and $v_{32} (n_2) = L_0^{32,\kappa} + n_2 L_{a^{\parallel}}^{\kappa}$ depend on the configuration; $L_0^{21,\kappa}$ is the distance between the intersections $1$ and $2$ etc; $x=(z_{12}z_{34}/z_{13}z_{24})$ is the only actually independent coordinate; and the various hypergeometric functions are given by:
\begin{eqnarray}
F_1^{\kappa}(1-x) &\equiv& F(\nk, \lk, \nk + \lk, 1-x) \nonumber \\
F_2^{\kappa}(1-x) &\equiv& F(1-\nk, 1-\lk, 2-\nk - \lk, 1-x) \nonumber \\
K_1^{\kappa}(x) &\equiv& F(\nk, \lk, 1, x) \nonumber \\
K_2^{\kappa}(x) &\equiv& F(1-\nk, 1-\lk, 1, x) = (1-x)^{\nk + \lk - 1} K_1(x) \nonumber \\
\tau^{\kappa}(x) &\equiv& (1-x)^{1-\nk - \lk} \frac{B(1-\nu,1-\lk)}{B(\nk, 1-\nk)} \frac{F_2(1-x)}{K_1 (x)} = \frac{B(1-\nu,1-\lk)}{B(\nk, 1-\nk)} \frac{F_2(1-x)}{K_2 (x)}\nonumber \\
\tau^{\kappa \prime}(x) &\equiv& \frac{B(\nk,\lk)}{B(\nk,1-\nk)} \frac{F_1(1-x)}{K_1 (x)} \nonumber \\
I^{\kappa}(x) &\equiv& (1-x)^{1-\nk -\lk} B(\nk, 1-\nk) K_1^{\kappa} (x) K_2^{\kappa} (x) (\tau^{\kappa} + \tau^{\kappa \prime})
\end{eqnarray}
The manifestly-$SL(2,\mathcal{C})$-invariant form of the above is obtained by premultiplying the amplitude by $(z_4)^{2h_{\lambda}}$ and taking $\{z_1,z_2,z_3,z_4\} \rightarrow \{0,x,1,\infty\}$. To link the above expression with that in \cite{Cvetic:Yukawa}, we must put $d_2^{\kappa} = v_{32}^{\kappa}$, $d_1^{\kappa} = d_2^{\kappa} + \beta v_{21}$, where 
\begin{equation}
\beta^{\kappa} = \tau^{\kappa\prime} - \tau^{\kappa} = \frac{\sin \pi (1-\nk - \lk)}{\sin \pi \lk}
\end{equation}
In the limit considered there, they set $d_2^{\kappa} = d_1^{\kappa} \tau/\tau^{\prime}$. However, we continue with the above expression, and first normalise it by considering the limit $x \rightarrow 0$, i.e. $z_2 \rightarrow z_1$, $z_4 \rightarrow z_3$. This gives
\begin{equation}
\mathcal{A}_{tree} (x \rightarrow 0) \sim (\alpha^{\prime})^{-1} g_o^2 \bra \sigma^{(aa^{\parallel})}_0 (0) \sigma^{(a^{\parallel}a)}_0 (1) \ket_a \prod_{\kappa=1}^{3} C^{(aba^{\parallel})0}_{\nk,1-\nk} C^{(a^{\parallel}ca)0}_{1-\lk,\lk}  
\end{equation}
where we now have \emph{stretch} fields for strings stretched between parallel branes. In the effective field theory, these are highly massive for large separation, and thus non-supersymmetric; we normalise them for consitency with the case $a^{\parallel} = a$ to give $\bra 1 \ket_a$, in which case we obtain
\begin{equation}
\mathcal{A}_{tree} (x \rightarrow 0) \sim  (\alpha^{\prime})^{-2} g_a^{2} G_{C_{ab}\bar{C}_{ba}} G_{C_{ac}\bar{C}_{ca}}
\end{equation}  
Applying this to our expression, in this limit $\tau^{\kappa \prime} \rightarrow \tau \rightarrow \frac{-\sin \pi \nk}{\pi} \ln x$, and so we the instanton sum over $n_2$ vanishes - requiring us to poisson resum as in \cite{Cvetic:Yukawa}, giving
\begin{equation}
\mathcal{A}_{tree} (x \rightarrow 0) \sim C (\alpha^{\prime})^{-1} g_o^2 \prod_{\kappa=1}^3 z_{12}^{-2h_{\nu^{\kappa}}} z_{34}^{-2h_{\lambda^{\kappa}}} \frac{\sqrt{2\pi\alpha^{\prime}}}{L_a^{\kappa}} x^{\frac{(y^{\kappa})^2}{4\pi^2\alpha^{\prime}}}
\end{equation}
where $y^{\kappa}$ is the perpendicular distance between branes $a$ and $a^{\parallel}$ in sub-torus $\kappa$; the zero-mode of $v_{32}$ is irrelevant. This gives us two pieces of information: first, we obtain the normalisation
\begin{equation}
C = \frac{e^{\Phi}}{\alpha^{\prime}g_o^{2}} G_{C_{ab}\bar{C}_{ba}} G_{C_{ac}\bar{C}_{ca}} (2\pi)^{5/2}
\end{equation}
(almost) in agreement with the similar case considered by \cite{Cvetic:Yukawa}. Secondly, we obtain the conformal weight of the operators $\sigma^{aa^{\parallel}}$: $h_{aa^{\parallel}}^{\kappa} = \frac{(Y^{\kappa})^2}{2\pi^2\alpha^{\prime}}$.

Now we must determine the more general coefficients by taking the limit $z_3 \rightarrow z_2$, or equivalently $x \rightarrow 1$. In this case $\tau^{\kappa} \rightarrow 0$, $\tau^{\kappa \prime} \rightarrow \beta$, and we obtain
\begin{equation}
\mathcal{A}_{tree}(x \rightarrow 1) \sim C (\alpha^{\prime})^{-1} g_o^2 \prod_{\kappa=1}^3 (1-x)^{-\nk \lk} Y^{\kappa}  \sum_{n_1,n_2} \exp -\frac{A(n_1,n_2)}{2\pi\alpha^{\prime}} 
\end{equation}
where 
\begin{equation}
Y^{\kappa} = \left\{ \begin{array}{cc}  \bigg(\frac{\Gamma(1-\nk) \Gamma(1-\lk) \Gamma(\nk + \lk)}{\Gamma(\lk)\Gamma(\nk) \Gamma(1 - \nk - \lk)} \bigg)^{1/2}\qquad & \nk + \lk < 1 \\
\bigg(\frac{\Gamma(\nk) \Gamma(\lk)\Gamma(2-\nk-\lk)}{\Gamma(1-\nk)\Gamma(1-\lk) \Gamma(\nk + \lk - 1)} \bigg)^{1/2} \qquad & \nk + \lk > 1 \end{array}\right. 
\end{equation}
and $A^{\kappa}(n_1^{\kappa},n_2^{\kappa})$ is the sum of the areas of the two possible Yukawa triangles formed by the intersection of the three branes wrapping in both directions, given by
\begin{eqnarray}
A^{\kappa}(n_1^{\kappa},n_2^{\kappa}) &=& \frac{1}{2} \frac{\sin \pi \nk \sin \pi \lk}{\sin \pi (1-\nk - \lk)} \left( (v_{32}^{\kappa})^2 + (v_{21}^{\kappa} \beta^{\kappa} + v_{32}^{\kappa})^2\right) \nonumber \\
&=& \frac{1}{2} \frac{\sin \pi \nk \sin \pi \lk}{\sin \pi (1-\nk - \lk)} \left( (d_1^{\kappa})^2(n_1^{\kappa}) + (d_2^{\kappa})^2 (n_2^{\kappa}) \right)
\end{eqnarray}
(and a similar expression for $\lk + \nk > 1$). With this expression, it becomes immediately clear how we can obtain the OPE coefficients:
\begin{equation}
\mathcal{A}_{tree}(x \rightarrow 1) \sim (\alpha^{\prime})^{-1} g_o^{-2} \bra 1 \ket_c g_0^4 \prod_{\kappa=1}^3 C_{1-\nk,1-\lk}^{(ba^{\parallel}c)1-\nk-\lk} C_{\nk,\lk}^{(cab)\nk + \lk} C^{(cbc)0}_{\lk,1-\lk} (1-x)^{-\nk\lk}
\end{equation}
and thus we infer
\begin{multline}
\sqrt{\alpha^{\prime}}g_o e^{-\Phi/2} \prod_{\kappa=1}^3 C_{\nk,\lk}^{(cab)\nk + \lk} = \left( 2\pi G^{C_{bc}\bar{C}_{cb}} G_{C_{ab}\bar{C}_{ba}} G_{C_{ac}\bar{C}_{ca}}\right)^{1/2} \\
\prod_{\kappa=1}^3 (\sqrt{2\pi}Y^{\kappa})^{1/2} \sum_{m^{\kappa}} \exp -\frac{A^{\kappa}(m^{\kappa})}{2\pi\alpha^{\prime}}  
\label{abcOPE}\end{multline}
where now $A^{\kappa}(m^{\kappa})$ is the area of the triangle $bac$, while the conjugate coefficient will have the area of $ba^{\parallel}c$ (zero if $a^{\parallel}=a$). This concludes the determination of all the relevant OPE coefficients in the theory. For use in the text, we define
\begin{equation}
Y^{(cab)} = g_o \prod_{\kappa=1}^3 C^{(cab)}(G_{C_{bc}\bar{C}_{cb}})^{1/2}
\label{YukDef}\end{equation}
and $Y^{ba^{\parallel}c}$ analagously; they represent the physical yukawa couplings.

\section{One-Loop Amplitudes With Gauge Bosons}\label{LoopTech}

\subsection{Classical Part}

The prescription of \cite{Atick:Twist} applies to the evaluation of the classical part of boundary-changing operators, by the analogy with twist operators, after we have applied the doubling trick - the net result being that we must halve the action that we obtain. Since it is the only consistent arrangement for two and three point diagrams, and the most interesting for four-point and above, we shall specialise to all operators lying on the imaginary axis, where the domain of the torus (doubled annulus) is $[-1/2,1/2]\times [0,it]$; this provides simplifications in the calculation while providing the most interesting result. In the case of all operators on the other end of the string, we would simply define the doubling differently to obtain the same result, and since the amplitude only depends on differences between positions the quantum formulae given would be correct. 

For $L$ operators inserted, we must have $M = \sum_{i=1}^{L} \theta_i $ integer to have a non-zero amplitude; for vertices chosen to lie on the interior of a polygon we will have $M = L - 2$, whic h shall be the case for the three and four point amplitudes we consider - in the case of a two-point amplitude we must have $M=1$. Labelling the vertices in order, we denote the first $L-M$ vertices $\{z_{\alpha}\}$, and the remaining $M$ vertices $\{z_{\beta}\}$, we then have a basis of $L-M$ functions for $\partial X (z)$ given by

\begin{equation}
\omega^{\alpha} = \theta_1 (z - z_{\alpha} - Y) \prod_{j \in \{\alpha\}\ne \alpha}^{L-M} \theta_1 (z-z_{j}) \prod_{k=1}^{L} \theta_1(z-z_k)^{\theta_k - 1}
\end{equation}
where
\begin{equation}
Y = -\sum_{i=1}^{L-M} \theta_i \ z_i  + \sum_{j=L-M+1}^{L} (1-\theta_j ) z_j
\end{equation}
and the basis of $M$ functions for $\partial \bar{X} (z)$
\begin{equation}
\omega^{\beta} = \theta_1 (z - z_{\beta} + Y) \prod_{j \in \{\beta\} \ne \beta}^{L} \theta_1 (z-z_{j}) \prod_{k=1}^{L} \theta_1(z-z_k)^{-\theta_k}
\end{equation}
We are then required to choose a basis of closed loops on the surface. There are two cycles associated with the surface which we shall label $A$ and $B$, and define as follows:
\begin{eqnarray}
\oint_{\gamma_A}  \d z &=& \int_{it - 1/2}^{-1/2} \d z \nonumber \\
\oint_{\gamma_B} \d z &=& \int_{-1/2}^{1/2} \d z
\end{eqnarray}
The remaining integrals involve the boundary operators, and we define (not-closed) loops $C_i$ by 
\begin{equation}
\int_{C_i} \d z  = \int_{z_i}^{z_{i+1}} \d z 
\end{equation}
Note that we have defined in total $L-1$ $C$-loops, and they are actually linearly dependent, since we can deform a contour around all of the operators to the boundary to give zero - we only require $L-2$. These could then be formed into (closed) pochhamer loops by multiplication by a phase factor, but it actually turns out that this is not necessary. We form the above into a set $\{\gamma_a\} = \{\gamma_A, \gamma_B\} \cup \{C_i,i=1,...,L-2\}$, and define the $L\times L$ matrix $W^i_a$ by
\begin{eqnarray}
W^{\alpha}_a &\equiv&  \int_{\gamma_a} \d z \omega^{\alpha} (z) \nonumber \\
W^{\beta}_a &\equiv& \int_{\bar{\gamma}_a}\d \bar{z} \bar{\omega}^{\beta} (\bar{z})
\end{eqnarray}

The boundary operators induce branch cuts on the worldsheet, which we have a certain amount of freedom in arranging. We shall choose the prescription that the cuts run in a daisy-chain between the operators, with phases $\exp(i\alpha_i)$ when passing through the cut anticlockwise with respect to $z_i$ defined as follows:
\begin{eqnarray}
\alpha_1 &=& 2 \pi \theta_1 \nonumber \\
\alpha_{L-1} &=& -2 \pi \theta_L \nonumber \\
2 \pi \theta_i &=& \alpha_i - \alpha_{i-1} \nonumber \\
\alpha_i &=& 2\pi \sum_{j=1}^{i} \theta_j
\end{eqnarray}
noting that $\alpha_i$ is only defined modulo $2\pi$. Each path $\gamma_{a}$ is associated with a physical displacement $v_a$ (which shall be determined later); i.e  
\begin{equation}
\Delta_{\gamma_a} X_{cl} = v_a
\end{equation}   
and, since we can write $\partial X$ and $\partial \bar{X}$ as linear combinations of $\omega^{\alpha}$ and $\omega^{\beta}$ respectively, we obtain
\begin{equation}
S = \frac{1}{4\pi \alpha^{\prime}} v_a \bar{v}_b \{ (W^{-1})^a_{i^{\prime}} (\bar{W}^{-1})^b_{j^{\prime}} (\omega^{i^{\prime}}, \omega^{j^{\prime}} ) + (W^{-1})^a_{i^{\prime\prime}} (\bar{W}^{-1})^b_{j^{\prime\prime}} (\omega^{j^{\prime\prime}}, \omega^{i^{\prime\prime}}) \}
\label{FullAction}\end{equation}
where the sum over primed indeces is understood to be over $\{\alpha\}$, and over double-primed indeces to be over $\{\beta\}$, and for reference we have used the complexification $X = \frac{1}{\sqrt{2}}(X_1 + i X_2)$. The inner product is defined to be
\begin{equation}
(\omega^{i^{\prime}}, \omega^{j^{\prime}}) = \int \d^2 z \omega^{i^{\prime}} (z) \bar{\omega}^{j^{\prime}} (\bar{z}) \equiv i W^{i^{\prime}}_a \bar{W}^{j^{\prime}}_b M^{ab}
\end{equation}
and similarly
\begin{equation}
(\omega^{i^{\prime\prime}}, \omega^{j^{\prime\prime}}) = \frac{1}{4 \pi \alpha^{\prime}} \int \d^2 z \omega^{i^{\prime\prime}} (z) \bar{\omega}^{j^{\prime\prime}} (\bar{z}) \equiv i \bar{W}^{i^{\prime\prime}}_a W^{j^{\prime\prime}}_b \bar{M}^{ab}
\end{equation}
and thus $\bar{M}^{ab} = -M^{ba}$. As in \cite{Atick:Twist} we have $M^{AB} = -M^{BA} = 1$, but, after performing the canonical dissection on the torus for our arrangement of branch cuts and basis loops we determine:
\begin{equation}
\begin{array}{cccc}M^{ml} &=&\frac{2i}{\sin \big(\frac{\alpha_{L-1}}{2}\big)} e^{ i \frac{\alpha_l - \alpha_m}{2}} \sin\bigg(\frac{\alpha_{L-1} - \alpha_l}{2}\bigg) \sin \frac{\alpha_m}{2} & m < l \\
M^{mm} &=&\frac{2i}{\sin \big(\frac{\alpha_{L-1}}{2}\big)} \sin\bigg(\frac{\alpha_{L-1} - \alpha_m}{2}\bigg) \sin \frac{\alpha_m}{2} &\end{array}
\label{M}\end{equation}
Inserted into the expression for the action, this gives $S = \frac{i}{4\pi \alpha^{\prime}} v_a \bar{v}_b S^{ab}$, where 
\begin{equation}
S^{ab} = (\bar{W}^{-1})^b_{j^{\prime}} \bar{W}^{j^{\prime}}_d M^{ad} + (W^{-1})^a_{i^{\prime\prime}} W^{i^{\prime\prime}}_d \bar{M}^{bd}
\end{equation}

At this point we determine the displacements. Clearly, since the boundary $Re(z)=-1/2$ contains no boundary-changing operators, along $\gamma_A$ the string end is fixed to one brane, which we shall label $a$. Thus, $v_A$ represents the one-cycles of brane $a$, and is equal to $\frac{1}{\sqrt{2}}n_A L_a$, where $L_a$ is the length of $a$ and the factor of $1/\sqrt{2}$ is due to the complexification we chose. Now, the technique that we are using requires there to be a path around the boundary of the worldsheet where we do not cross any branch cuts: if the string has one end fixed on brane $a$ along path $\gamma_A$, the other end (at $z=0$) must either reside on brane $a$ or a brane parallel to it, which we shall denote $a^{\parallel}$. Path $\gamma_B$ is related to the displacement between these branes. Consider the doubling used, for coordinates aligned along brane $a$:
\begin{equation}
\partial X (z) = \left\{ \begin{array}{cc} \partial X (z) & \Re(z) > 0 \\
-\bar{\partial}  \bar{X} (-\bar{z}) & \Re(z) <0  \\ \end{array} \right.
\end{equation}
and a similar relationship for $-\bar{\partial} X(-\bar{w})$ and $\partial \bar{X}$. Note that we have $\Delta_{\gamma_A}X = \Delta_{\gamma_a}\bar{X}$, in keeping with our identification of $v_A$. We have  
\begin{eqnarray}
\Delta_{\gamma_B} &=& \int_{\gamma_B} \d z \partial X + \int_{\gamma_B} \d \bar{z} \bar{\partial} X (z) \nonumber\\
&=& \int_{0}^{1/2} \d x \frac{\d}{\d x} X (x) - \frac{\d}{\d x} \bar{X} (x)\nonumber \\
&=& i\sqrt{2} (X_2 (a) - X_2 (a^{\parallel}) ) \nonumber \\
&=& i\sqrt{2} (\frac{n_B 4\pi^2 T_2}{L_a} + y)
\end{eqnarray}
where $y$ is the smallest distance between $a$ and $a^{\parallel}$ and $T_2$ is the K\"ahler modulus of the torus, defined earlier. This resolves an ambiguity in \cite{Ben}. The remaining paths have straightforward identifications, since they are essentially like the tree-level case; they are the displacements between physical vertices. We must identify each portion of the $Re(z)=0$ boundary with a brane segment between intersections; so for the case
\begin{equation}
\bra \sigma_{\nu}^{(ab)} (z_1) \sigma_{1-\nu}^{(ba^{\parallel})} (z_2)\sigma_{1-\lambda}^{(a^{\parallel}c)} (z_3) \sigma_{\lambda}^{(ca)} (z_4) \ket_{aa}
\end{equation}
we have $v_1 = \frac{1}{\sqrt{2}}( n_1 L_b + L_{12})$ and $v_2 = \frac{1}{\sqrt{2}}(n_2 L_{a^{\parallel}} + L_{23})$. If we were to use Pochamer loops for these, we would multiply these by Pochamer factors - but these would be cancelled in the action by those associated with the loops on the worldsheet. 

Note that although we are free to choose the arrangement of branch cuts on the worldsheet, we have no freedom in identifying the displacement vectors, and thus the daisy-chain prescription is the simplest, particularly since we are restricted in the permutations of the boundary-changing operators - we must ensure that each change of brane has the correct operator to mediate it.

\subsection{Quantum Part}

The quantum part of the correlator is given in terms of the variables defined in the previous section as
\begin{multline}
\bra \prod_{n=1}^{L} \sigma_{\theta_n} (z_n) \ket = |W|^{-1/2} \theta_1 (Y)^{L-2} \prod_{i<j}^{L-M} \theta_1 (z_i - z_j) \prod_{L-M<i<j}^L \theta_1(z_{i} - z_{j})\\
\prod_{i < j}^L \theta_1 (z_i - z_j)^{-\theta_i \theta_j -(1-\theta_i)(1-\theta_j)}
\end{multline}
where $|W^{\kappa}|$ is the determinant of the matrix of integrals of cut differentials. Note that this expression does not depend upon the brane labels of the boundary-changing operators, as it is not sensitive to the particular boundary conditions.

The correlators for the fermionic component of the vertex operators do not receive worldsheet instanton corrections, so they are given by the relatively simple formula presented in \cite{Ben}:
\begin{equation}
\bra \prod_{i=1}^{L} e^{iq_i H (z_i)} \ket_{\nu} = \theta_{\nu} (\sum_{i=1}^{L} q_i z_i ) \prod_{i < j} \theta_1 (z_i - z_j)^{q_i q_j} 
\label{spinops}\end{equation}
the above is a generalisation of the formulae for standard spin-operator correlators (see e.g. \cite{AtickDixonSen}) and is thus also valid for the $4d$ spin correlators.

\subsection{4d Bosonic Correlators}

Beginning with the Green's function for the annulus, obtained via the method of images on the torus:
\begin{multline}
G (z,w) = -\frac{\alpha^{\prime}}{2} \ln |\theta_1(z-w)|^2 -\frac{\pi i \alpha^{\prime}}{t} (\Im (z-w))^2 \\
-\frac{\alpha^{\prime}}{2} \ln |\theta_1(z+\bar{w})|^2 -\frac{\pi i \alpha^{\prime}}{t} (\Im (z+\bar{w}))^2
\end{multline}
and specialising to all points at boundaries (i.e. $\bar{z} = -z$ or $1-z$), we obtain the amplitude on the annulus
\begin{eqnarray}
A^4_X \equiv \langle \prod_{i=1}^{M} e^{i k_i \cdot X(z_i)}\rangle &=& C_X \prod_{i<j} \bigg[ \th{i}.{j} e^{\frac{-\pi}{t} (\Im(z_i - z_j))^2} \bigg]^{2\alpha^{\prime} k_i\cdot k_j} \nonumber \\
&=& Z_X^4 \prod_{i<j} \bigg[ \frac{\th{i}.{j}}{\theta_{1}^{\prime} (0)} e^{\frac{-\pi}{t} (\Im(z_i - z_j))^2} \bigg]^{2\alpha^{\prime} k_i\cdot k_j} \nonumber \\
&\equiv& Z_X^4 \prod_{i<j} \chi^{2\alpha^{\prime} k_i \cdot k_j}
\label{A4X}\end{eqnarray}
where 
\begin{equation}
Z_X^4 = (8\pi^2 \alpha^{\prime} t)^{-2} \eta (it)^{-2}
\end{equation}
is the partition function for the non-compact bosons with the $bc$-ghost contribution included. We then use the green's function to obtain
\begin{multline}
\langle \partial X^{\mu_3}(z) \partial X^{\mu_4}(w) \prod_i e^{i k_i \cdot X(z_i)} \rangle = \langle \prod_{m=1}^{4} e^{i k_m \cdot X(z_m)}\rangle\\
\bigg\{ 2\eta^{\mu \nu} \partial_z \partial_w G(z,w) +4\sum_{i=1}^4 \sum_{j=1}^4 k_i^{\mu_3} k_j^{\mu_4} \partial_z G(z,z_i) \partial_w G(w,z_j) \bigg\} 
\end{multline}
which is required for 4-point functions.

\section{One-Loop Amplitudes Without Gauge Bosons}\label{AdvancedLoopTech}

\subsection{Classical Part}

\subsubsection{Cut Differentials}

The cut differentials for diagrams on the doubled annulus where there exists no periodic cycle necessarily have modified boundary conditions. Where the diagram with no boundary changing operators is the partition function of strings streched between branes $a$ and $b$ with angle $\pi\theta_{ab}$, the conditions for the one form $\partial X(w)$ are
\begin{eqnarray}
\partial X (w+it) = \partial X (w) \nonumber \\
\partial X(w+1) = e^{2\pi i \theta_{ab}} \partial X (w)
\end{eqnarray}
while 
\begin{eqnarray}
\partial \bar{X} (w+it) = \partial X (w) \nonumber \\
\partial \bar{X}(w+1) = e^{-2\pi i \theta_{ab}} \partial X (w)
\end{eqnarray}
while the local monodromies remain the same as for the periodic case. Indeed, we can retain many of the elements of those differentials, including
\begin{eqnarray}
\gamma_X (z) &=& \prod_{i=1}^L \theta_1 (z-z_i)^{\theta_i - 1} \nonumber \\
\gamma_{\bar{X}} (z) &=& \prod_{i=1}^L \theta_1 (z-z_i)^{-\theta_i}
\end{eqnarray}
These satisfy the local monodromies; to construct a complete set of differentials satisfying the global monodromy conditions, we note the identities:
\begin{eqnarray}
\theta \left[ \begin{array}{c} 1/2 + a \\ 1/2 \end{array} \right] (z+m;\tau) &=& \exp (2\pi i (1/2+a)m) \theta \left[ \begin{array}{c} 1/2 + a \\ 1/2 \end{array} \right] (z;\tau)  \\
\theta \left[ \begin{array}{c} 1/2 + a\\ 1/2 \end{array} \right] (z+m\tau;\tau) &=& \exp (-2\pi i m/2) \exp (-\pi i m^2 \tau - 2 \pi i mz)\theta \left[ \begin{array}{c} 1/2 +a \\ 1/2 \end{array} \right] (z;\tau) \nonumber
\end{eqnarray}
which show that we only need to modify one theta function from the periodic case; we have also
\begin{equation}
\theta \left[ \begin{array}{c} c + a \\ b \end{array} \right] (z;\tau) = \exp[2\pi i a (z+c) +a^2 \pi i \tau] \theta \left[ \begin{array}{c} c \\ b \end{array} \right] (z + a\tau;\tau)
\end{equation}
which is crucial for showing the equivalence between the approach that we are about to use, and the method of obtaining these amplitudes by factorising higher-order amplitudes calculated by the previous method. Denoting the theta-function
\begin{equation}
\theta_{\pm ab} (z,\tau) \equiv \theta \left[ \begin{array}{c} 1/2 \pm \theta_{ab} \\ 1/2 \end{array} \right] (z;\tau)
\end{equation}
we construct the set of $L-M$ differentials for $\partial X(z)$, similar to before 
\begin{equation}
\omega^{\alpha}_{+ab} (z) = \gamma_X (z) \theta_{+ab} (z-z_{\alpha} - Y) \prod_{j \in \{\alpha\}\ne \alpha}^{L-M} \theta_1 (z-z_{j})
\label{TwistOmegaa}\end{equation}
where $Y$ is as defined before; and we have the set of $M$ differentials for $\partial \bar{X}$
\begin{equation}
\omega^{\beta}_{-ab} (z) = \gamma_{\bar{X}} (z) \theta_{-ab} (z-z_{\beta} + Y) \prod_{j \in \{\beta\}\ne \beta}^{L} \theta_1 (z-z_{j})
\label{TwistOmegab}\end{equation}
We demonstrate that these are complete sets in the same way as in \cite{Atick:Twist}: first, note that the function are independent, since $\omega^{\alpha}(z_{j \ne \alpha}) = 0$. Then suppose we had another differential $\omega^{\prime} (z)$; we construct the doubly-periodic meromorphic function $\lambda(z)$
\begin{equation}
\lambda (z) = \frac{\omega^{\prime}(z)}{\omega^{1} (z)} - \sum_{i=1}^{L-M} C_i \frac{\omega^{i}(z)}{\omega^1 (z)}
\end{equation}
At $z=z_1$ or $z \in \{z_{\beta}\}$, the above is not singular, while we can adjust the $L-M$ constants $C_i$ to cancel the residues of the poles at $z=z_{i \ne 1}$ and $z_1 + Y-\theta_{ab}it$. Thus, since $\lambda$ has no poles, and it is doubly periodic on the torus, it is a constant(the last point contains the only subtlety with respect to the earlier case; even though the differentials $\omega^{\alpha}$ are not periodic on the torus, because they only acquire a phase, the differential $\lambda$ is periodic). Hence, since $C_1$ just multiplies a constant, we can adjust it to set $\lambda$ to zero. The same follows for $\omega^{\beta}$.

Note that if we want to use the same theta-functions throughout, we could choose a basis obtained by factorising the functions used earlier. In this case, we obtain
\begin{equation}
\omega^{\alpha\prime}_{+ab} (z) = e^{\pi i (1-it)\theta_{ab}} e^{2\pi i \theta_{ab} z} \gamma_X (z) \theta_{1} (z-z_{\alpha} - Y + \theta_{ab}it) \prod_{j \in \{\alpha\}\ne \alpha}^{L-M} \theta_1 (z-z_{j})
\end{equation}
and
\begin{equation}
\omega^{\beta\prime}_{-ab} (z) = e^{-\pi i (1-it)\theta_{ab}} e^{-2\pi i \theta_{ab} z} \gamma_{\bar{X}} (z) \theta_{1} (z-z_{\beta} + Y - \theta_{ab}it) \prod_{j \in \{\beta\}\ne \beta}^{L} \theta_1 (z-z_{j})
\end{equation}
The relation between the bases is given simply by
\begin{eqnarray}
\omega^{\alpha\prime}_{+ab} (z) &=& e^{\pi t(\theta_{ab}+\theta_{ab}^2)} e^{2\pi i \theta_{ab} (z_\alpha + Y)}\omega^{\alpha}_{+ab} (z) \nonumber \\
\omega^{\beta\prime}_{-ab} (z) &=& e^{-\pi t(\theta_{ab}+\theta_{ab}^2)} e^{-2\pi i \theta_{ab} (z_\beta - Y)}\omega^{\beta}_{-ab} (z)
\label{PrimeNoPrime}\end{eqnarray}

The choice of basis is not important for the following section, and the amplitude will of course be independent of the basis choice.

\subsubsection{Canonical Disection}

With our basis of cut differentials for the doubled annulus we require their hermitian inner product to calculate the action. This is given by
\begin{equation}
(\omega^i,\omega^j)=\int_{R} \d^2 z \bar{\omega}^j \omega^i = i \oint_{\partial R} \bar{\omega}^{j}(\bar{z})\d \bar{z} \int_{z_0}^{z} \d z \omega^i
\end{equation}
This is the canonical dissection of the surface, where the contour passes anticlockwise around the surface without crossing any branch cuts. The task is then to choose the most convenient arrangement of cuts, and express the above in terms of integrals of paths on the surface corresponding to physical displacements. Since we consider operators only on one boundary, we arrange them to be on the imaginary axis; since we always have one vertex fixed, we choose this to be $z_L$ and place it at the origin. The integrals between vertices are then defined by
\begin{equation}
\int_{C_n} = \int_{z_{n+1}}^{z_{n}}
\end{equation}
where $\Im(z_n) > \Im(z_{n+1})$. We also have the loops
\begin{eqnarray}
\int_{A} &=& \int_{-1+it}^{-1} = \int_{-1/2+it}^{-1/2} \nonumber \\
\int_{B} &=& \int_{-1}^{0}
\end{eqnarray}
and the phases
\begin{eqnarray}
\alpha_1 &=& 2 \pi \theta_1 +2 \pi \theta_{ab}\nonumber \\
\alpha_{L-1} &=& -2 \pi \theta_L + 2 \pi \theta_{ab}\nonumber \\
2 \pi \theta_i &=& \alpha_i - \alpha_{i-1} \nonumber \\
\alpha_i &=& 2\pi \sum_{j=1}^{i} \theta_j + 2\pi \theta_{ab}
\end{eqnarray}
we also label the additional cycle $\gamma_D$:
\begin{eqnarray}
\int_{\gamma_D} &=& \int_{z_1}^{it} \nonumber \\
\alpha_D &=& -2\pi\theta_{ab}
\end{eqnarray}
which we eliminate, along with $C_1$, via the equations
\begin{eqnarray}
\gamma_D + C_1 + \sum_{n=2}^{L-1} C_n &=& -\gamma_A \nonumber \\
e^{i\alpha_D} \gamma_D + e^{i\alpha_1} C_1 + \sum_{n=2}^{L-1} e^{i\alpha_n} C_n &=& -\gamma_A
\label{CycleSums}\end{eqnarray}
Hence we have chosen our set of paths to be $\{\gamma_A,\gamma_B\}\cup\{C_2..C_{L-1}\}$. With these definitions, we now perform the canonical dissection, and defining our matrices as before $W^{i^{\prime}}_A = \int_A \omega^{i^{\prime}}$ etc, the inner product is of the form
\begin{equation}
(\omega^{i^{\prime}}, \omega^{j^{\prime}}) = i W^{i^{\prime}}_a \bar{W}^{j^{\prime}}_b M^{ab}
\end{equation}
and the results are
\begin{eqnarray}
M^{AB} &=& 1 \nonumber \\
M^{AA} &=& 2i \frac{\sin \frac{\alpha_1}{2} \sin \frac{\alpha_D}{2}}{\sin \frac{\alpha_D - \alpha_1}{2}}\nonumber \\
M^{An} &=& -2i \frac{e^{-i\alpha_n}}{\sin(\frac{\alpha_D -\alpha_1}{2})} \sin (\frac{\alpha_n - \alpha_1}{2}) \sin \frac{\alpha_D}{2}\nonumber \\
M^{mn} &=& \frac{2i}{\sin (\frac{\alpha_D - \alpha_1}{2})} e^{i\frac{\alpha_m-\alpha_n)}{2}} \sin (\frac{\alpha_m -\alpha_D}{2}) \sin (\frac{\alpha_n - \alpha_1}{2})
\end{eqnarray}
where $m \ge n$ in the last line, $L-1 \ge n, m \ge 2$ and the elements reflected in the diagonal can be obtained from the above using $M^{dc} = -\bar{M}^{cd}$. We see that we can obtain appropriate expressions for the previous case simply by taking $\theta_{ab} =0$, in which case the formulae are greatly simplified, with the $AA$ and $An$-elements vanishing. The above expression then yields the classical action by equation (\ref{FullAction}).

\subsection{Quantum Part}

As for the classical part, the quantum part of the bosonic correlator for these diagrams may be determined in two ways; factorisation of a diagram with a larger number of operators inserted, or by calculating new green's functions and proceeding by the stress-tensor method. However, unlike for the classical part, it is straightforward to perform the factorisation. 

To obtain the quantum amplitude for $L$ boundary-changing operators with angles $\{\theta_{i}\}$ and overall periodicity $\theta_{ab}$, we start with an amplitude with $L+1$ operators with angles $\{\theta_{ab},\theta_i,\theta_{L}-\theta_{ab}\}$ which we assign to vertices at $\{z_0,z_i,0\}$. The above choice of angles ensures that in both sets of equations $M$ is the same, and we have
\begin{equation}
Y_{L+1} = Y_{L} -\theta_{ab}it
\end{equation}

The full quantum correlator is 
\begin{multline}
Z_{qu}=f(it)|W_{L+1}|^{-1/2} \theta_1 (Y_{L+1})^{(L-1)/2} \prod_{0\le i < j}^{L-M} \theta_1 (z_i - z_j)^{1/2} \prod_{L-M < i < j}^{L} \theta_1 (z_i - z_j)^{1/2} \\
\prod_{0 \le i < j}^{L} \theta_1 (z_i - z_j)^{-\frac{1}{2}[1 - \theta_i -\theta_j +2 \theta_i \theta_j]} 
\end{multline}
where $f(it)$ is the normalisation. We then note that in the limit $z_0 \rightarrow it$, all except two of the integrals in the matrix $W_{L+1}$ are finite. Indeed, only $\omega^{0}(z)$ develops a singularity, and only the integrals of it over the cycles $\gamma_B$ and $C_{0}$ become infinite (note that we are using the set of curves $\{\gamma_A,\gamma_B,C_0..C_{L-1}\}$). The determinant becomes in the limit
\begin{equation}
|W_{L+1}| \rightarrow  (W_{L+1})^0_{0} |W_{L}| - (W_{L+1})^0_B |W_{L}^{\prime}|
\end{equation}
where $|W_{L}^{\prime}|$ is the determinant with the $\gamma_B$ cycles and $\omega^0$ integrals deleted: we then note that the rows are not linearly independent due to the identities (\ref{CycleSums}), and it is therefore zero. We must finally evaluate $(W_{L+1})^0_0$. 
\begin{multline}
(W_{L+1})^0_0 \sim -i (it - z_0)^{\theta_L -1} B(1-\theta_L,\theta_L - \theta_{ab}) e^{\pi i (1-it)(\theta_{ab}-1)} \theta_1^{\prime}(0)^{\theta_L - 2} \\
\theta_1 (-Y_{L+1}) \prod_{i=1}^{L-M} \theta_1 (it-z_i)^{\theta_i} \prod_{j=L-M+1}^{L-1} \theta_1 (it-z_j)^{\theta_j-1}
\end{multline}
which, when we consider that the amplitude should factorise according to 
\begin{equation}
\sigma_{\theta_{ab}}^{(ca)} (z_0) \sigma_{\theta_{L}-\theta_{ab}}^{(ab)} (0)  \sim (it-z_0)^{-\theta_{ab}\theta_{L} + \theta_{ab}^2} C^{(bac)\theta_{L}}_{\theta_{L}-\theta_{ab},\theta_{ab}} \sigma^{(bc)}_{\theta_L} (0) 
\end{equation}
we find that the quantum portion of the amplitude is
\begin{multline}
Z_{qu}^{(ab)}= g(it) |W_{L}^{\prime}|^{-1/2} e^{2\pi i P}\theta_1 (Y_{L}-\theta_{ab}it)^{(L-2)/2} \prod_{0< i < j}^{L-M} \theta_1 (z_i - z_j)^{1/2}\\
 \prod_{L-M < i < j}^{L} \theta_1 (z_i - z_j)^{1/2} \prod_{0 < i < j}^{L} \theta_1 (z_i - z_j)^{-\frac{1}{2}[1 - \theta_i -\theta_j +2 \theta_i \theta_j]} 
\end{multline}
where now $W_{L}$ contains integrals of the primed basis of cut differentials, and 
\begin{equation}
P = Y + \frac{1}{2}\left(\sum_{i=1}^{L-M} z_i - \sum_{j=L-M+1}^{L} z_j \right)
\end{equation} 
Now we find that the ``natural'' basis for these functions is that which we defined in equations (\ref{TwistOmegaa}) and (\ref{TwistOmegab}); in this basis, using the relations (\ref{PrimeNoPrime}) we find the amplitude to be
\begin{multline}
Z_{qu}^{(ab)}= g(it) |W_{L}|^{-1/2} \theta_{-ab} (Y_{L})^{(L-2)/2} \prod_{0< i < j}^{L-M} \theta_1 (z_i - z_j)^{1/2}\\
 \prod_{L-M < i < j}^{L} \theta_1 (z_i - z_j)^{1/2} \prod_{0 < i < j}^{L} \theta_1 (z_i - z_j)^{-\frac{1}{2}[1 - \theta_i -\theta_j +2 \theta_i \theta_j]} 
\end{multline}

The above can also be obtained by repeating the analysis of \cite{Atick:Twist}; most of the steps are the same, since the quantum amplitude does not depend upon the exact form of the greens' functions, only certain constraints upon them, which remain the same for these amplitudes. 

\subsection{Fermionic Correlators}

The fermionic correlators for these amplitudes are easily obtained by simply factorising equation (\ref{spinops}); we obtain
\begin{multline}
\bra \prod_{i=1}^{L} e^{i(\theta_i - 1/2) H (z_i)} \ket_{\nu,\theta_{ab}} =  e^{2\pi i (\theta_{ab}-1/2)P} \theta_{\nu} (it(\theta_{ab}-1/2) + \sum_{i=1}^{L} q_i z_i) \\
\prod_{i < j} \theta_1 (z_i - z_j)^{(\theta_i - 1/2)(\theta_j - 1/2)} 
\end{multline}

\section{Theta Identities}\label{Thetas}

Throughout we use the standard notation for the Jacobi Theta functions:
\begin{equation}
\theta \left[ \begin{array}{c} a \\ b \end{array} \right] (z;\tau) = \sum_{n = -\infty}^{\infty} \mathrm{exp} \bigg[ \pi i (n+a)^2 \tau + 2 \pi i (n+a)(z+b) \bigg]
\end{equation}
and define $\theta_{\alpha \beta} \equiv \theta \left[ \begin{array}{c} \alpha/2 \\ \beta/2 \end{array} \right]$, so that we have the periodicity relations
\begin{eqnarray}
\theta \left[ \begin{array}{c} a \\ b \end{array} \right] (z+m;\tau) &=& \exp (2\pi i am) \theta \left[ \begin{array}{c} a \\ b \end{array} \right] (z;\tau)  \\
\theta \left[ \begin{array}{c} a \\ b \end{array} \right] (z+m\tau;\tau) &=& \exp (-2\pi i bm) \exp (-\pi i m^2 \tau - 2 \pi i mz)\theta \left[ \begin{array}{c} a \\ b \end{array} \right] (z;\tau) \nonumber
\end{eqnarray}
We also define, according to the usual conventions,
\begin{eqnarray}
\theta_1 \equiv \theta_{11} & \theta_2 \equiv \theta_{10} \nonumber \\
\theta_3 \equiv \theta_{00} & \theta_4 \equiv \theta_{01} 
\end{eqnarray}

\bibliography{ibwol}

\end{document}